\documentclass[11pt]{article}
\pdfoutput=1 
\usepackage{bbm,bm,graphicx,mathtools,color,slashed}
\usepackage{amssymb,amsmath}
\usepackage[colorlinks=true, pdfstartview=FitV, linkcolor=red, citecolor=blue, urlcolor=blue]{hyperref}
\usepackage[T1]{fontenc}
\usepackage[caption=false]{subfig}
\usepackage[nottoc]{tocbibind}
\usepackage[toc,page]{appendix}
\usepackage{authblk}
\usepackage[numbers,sort&compress]{natbib}
\usepackage{youngtab}
\usepackage[top=25truemm,bottom=25truemm,left=15truemm,right=15truemm]{geometry}
\usepackage{url}
\usepackage[margin=15truemm]{caption}
\usepackage{caption}
\usepackage{slashed}
\usepackage{ulem}
\numberwithin{equation}{section}
\allowdisplaybreaks

\usepackage{qcircuit}
\usepackage{mathtools}
\DeclarePairedDelimiter\bra{\langle}{\rvert}
\DeclarePairedDelimiter\ket{\lvert}{\rangle}
\DeclarePairedDelimiterX\braket[2]{\langle}{\rangle}{#1 \delimsize\vert #2}

\graphicspath{{./figures/}}

\newcommand{\diff}{\mathrm{d}}

\newcommand{\be}{\begin{equation}}      
\newcommand{\ee}{\end{equation}}      
\newcommand{\bea}{\begin{eqnarray}}      
\newcommand{\eea}{\end{eqnarray}}

\newcommand{\e}{\mathrm{e}}

\newcommand{\calO}{\mathcal{O}}

\newcommand{\calQ}{\mathcal{Q}}

\newcommand{\Tr}{\mathrm{Tr}}

\newlength{\fighskip} \fighskip=2pt
\newlength{\figvskip} \figvskip=3pt

\newcommand*{\figbox}[2]{{
 \def\figscale{#1}
 \def\arraystretch{0.8}
 \arraycolsep=0pt
 \begin{array}{c}
\vbox{\vskip\figscale\figvskip
  \hbox{\hskip\figscale\fighskip
    \includegraphics[scale=\figscale]{#2}}}
 \end{array}}}
\begin{document}
\title{Hayden-Preskill decoding from noisy Hawking radiation}

\author[1,2]{Ning Bao\thanks{ningbao75@gmail.com}}
\author[3]{Yuta Kikuchi\thanks{yuta.kikuchi@riken.jp}}

\affil[1]{\it Computational Science Initiative, Brookhaven National Laboratory, Upton, New York, 11973}
\affil[2]{\it Center for Theoretical Physics and Department of Physics, University of California, Berkeley}
\affil[3]{\it RIKEN BNL Research center, Brookhaven National Laboratory, Upton, NY, 11973, USA}

\date{}

\maketitle

\abstract
In the Hayden-Preskill thought experiment, the Hawking radiation emitted before a quantum state is thrown into the black hole is used along with the radiation collected later for the purpose of decoding the quantum state. 
A natural question is how the recoverability is affected if the stored early radiation is damaged or subject to decoherence, and/or the decoding protocol is imperfectly performed. 
We study the recoverability in the thought experiment in the presence of decoherence or noise in the storage of early radiation.

%\tableofcontents

%%%%%%%%%%%%%%%%%%%%%%%%%%%%%%%%%%%%%%%%%%%%%%%%%
\section{Introduction}
The Hayden-Preskill thought experiment \cite{Hayden:2007cs} asked the question of how much information could be recovered regarding a quantum state that was thrown into a black hole under the assumption that the black hole was a fast scrambling quantum system. 
The surprising result is that, given sufficient time after the message was thrown in, there existed a quantum channel for the recovery of the quantum information associated with the state that was thrown in (see also \cite{Abeyesinghe:2006,Hayden:2007}). This result was critical to the development of black hole quantum information as a subfield, in particular being critical to the formulation of the black hole information paradox \cite{Hawking:1976ra,Page:1993df,Page:1993wv,Almheiri:2012rt,Almheiri:2013hfa}. In more recent works, explicit quantum circuit implementations of the thought experiment have been proposed by \cite{Yoshida:2017non,Yoshida:2018vly,Yoshida:2018ybz}, some in the slightly modified context of traversable wormholes in AdS/CFT \cite{Gao_2017,Maldacena_2017}.

Of particular interest to those without a strong interest in quantum gravity, this protocol does not require that the quantum state even be thrown into specifically a black hole; indeed, all that is required is that it becomes associated with a sufficiently rapidly scrambling quantum system. Such understandings, in turn, have boosted the study of quantum chaos from the viewpoint of the quantum information theory and quantum many body physics~\cite{Sekino:2008he,Lashkari:2011yi,Hosur:2015ylk,Roberts:2016hpo,Nahum:2016muy,vonKeyserlingk:2017dyr,Cotler:2017jue,Khemani:2017nda,Yoshida:2019qqw,Kudler-Flam:2019wtv,Kudler-Flam:2020yml,Yan:2020fxu,Piroli:2020dlx}.
In this context, the work of \cite{Hayden:2007cs} could have potential applications for NISQ quantum algorithms and devices, perhaps as a subroutine or a kind of quantum memory. Studying this model in this context, however, naturally begs the question of robustness against decoherence.

Recently, the effects of certain types of noise and decoherence on the recoverability have been studied in~\cite{Yoshida:2018vly} to experimentally assess the amount of chaos in strongly correlated quantum systems on a noisy quantum device~\cite{Landsman:2018jpm}.
%They studied the effect of decoherence and noise on the Hayden-Preskill decoding protocol for the purpose of experimentally diagnosing chaos and information scrambling in quantum systems. 
While out-of-time ordered correlation (OTOC) functions are powerful diagnostic tools of information scrambling/chaos, their signals can be hard to extract due to decoherence and noise effects in realistic experimental setups. Yoshida and Yao~\cite{Yoshida:2018vly} proposed a Hayden-Preskill decoding protocol as an alternative tool to circumvent the difficulty. Accordingly, their primary interest lies in the effect of those errors on the operations of unitaries $U$, which can be noisy or imperfectly implemented.

% \red{
% On the other hand, the present study is focused on the effect of decoherence or noise in the storage of early radiation in line with the setup of Hayden-Preskill thought experiment. Namely, Bob has collected the early radiation from the black hole before Alice throws her state into it. However, Bob's collected radiation has been exposed to the interaction with environment by the time of decoding. Note that this is not necessarily the case in the experimental setup proposed in~\cite{Yoshida:2018vly}, where there is no temporal separation between the black hole evolution $U$ and backward evolution $U^\ast$ for recovery.}

In this work, we study the robustness of the Hayden-Preksill thought experiment to decohering effects different from those considered in \cite{Yoshida:2018vly}. The effect of decoherence on black hole quantum information has already been studied in for example \cite{Bao_2018,Agarwal:2019gjk}, and have been shown to have potentially quite interesting effects in the context of the information paradox. 
Here, however, we will focus on the purely information theoretic question of the robustness of the Hayden-Preskill recovery channel against erasure or decoherence, as could occur because of circuit imperfections or inadequate shielding against interactions with the environment. In this context, we assess the severity of those errors in terms of the quality of Hayden-Preskill decoding assuming that the erasure error or decoherence occurs in the stored radiation collected earlier for the purpose of recovery. 
After relating the recoverability to the mutual information between the input state and collected radiation in general setup, we carry out explicit computations of decoding fidelity, that is a proxy of recoverability, and the effects of errors provided the time evolution is described by a Haar random unitary.
We find that the erasure errors severely affects the decoding fidelity in contrast to decoherence, which has a weaker effect.

The organization of this work will be as follows. In Sec.~2, we will review the Hayden-Preskill thought experiment in the absence of decohering effects. In Sec~3, we will discuss the effect of erasure errors on the Hayden-Preskill protocol. In Sec~4, we will discuss the effects of decoherence. In Sec.~5, we compare the recoverabilities with the erasures and the decoherence based on the computations done in the previous sections. Finally, we will conclude with a discussion of these effects in section 5, in particular by contrast to the earlier work performed in \cite{Yoshida:2018vly}.
%%%%%%%%%%%%%%%%%%%%%%%%%%%%%%%%%%%%%%%%%%%%%%%%%
\section{Preliminary: Ideal Hayden-Preskill decoding}

%--------------------------------%
\subsection{Hayden-Preskill thought experiment}

Suppose Alice has a quantum diary represented by a state on $A$, which she throws into a black hole $B$. In particular, let the black hole have already emitted half of its qubits as Hawking radiation $B'$ when Alice throws her diary in, e.g. existing past its Page time~\cite{Page:1993df,Sekino:2008he}.
The composite system will then experience some unitary evolution $U$ to end up with Hawking radiation $D$ and the remainder of the black hole $C$.
A natural question to ask at this point would be if Bob can reconstruct Alice's diary by collecting the early radiation $B'$ and the late radiation $D$ and acting on it with some quantum recovery channel. If yes, a further question would be  how much late time radiation Bob needs in order to perform the reconstruction.
According to \cite{Hayden:2007cs}, the answer is affirmative, with Bob needing only to collect a small amount of the late time radiation to perform the recovery.

For the purpose of decoding, we attach a reference system $R$ that forms a EPR state with the state on $A$.
The state $\ket{\Psi_\text{HP}}$ after the time evolution $U$ is given by
\begin{align}
\label{eq:HPstate}
\hspace{-10mm} \ket{\Psi_\text{HP}} := (I_{R}\otimes U_{AB}\otimes I_{B'}) |\text{EPR}\rangle_{RA}\otimes |\text{EPR}\rangle_{BB'} = 
\figbox{0.4}{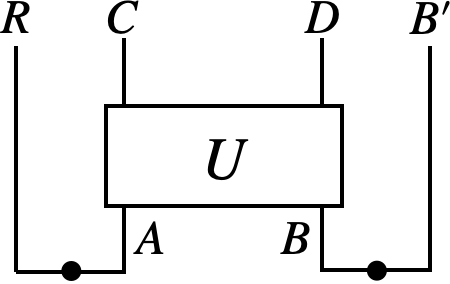} \qquad
\end{align}
The EPR states $\ket{\text{EPR}}_{RA}$ and $\ket{\text{EPR}}_{BB'}$ are defined by 
\begin{align}
\ket{\text{EPR}}_{RA} :=\frac{1}{\sqrt{d_A}}\sum_{i=1}^{d_A}\ket{i_R}\otimes\ket{i_A}
=\figbox{0.4}{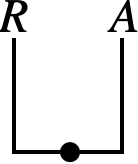},
\qquad
\ket{\text{EPR}}_{BB'} :=\frac{1}{\sqrt{d_B}}\sum_{i=1}^{d_B}\ket{i_B}\otimes\ket{i_{B'}}
=\figbox{0.4}{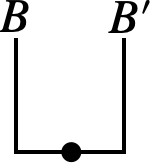},
\end{align}
with Hilbert space dimensions $d_A=d_R$ and $d_B=d_{B'}$. The dimension of the total Hilbert space is $d^2$ with 
$d=d_Ad_B=d_Rd_{B'}=d_Cd_D$.
The dot in the graphical representation stands for the normalization factor of a EPR state.

%Let us familiarize ourselves with the state representation of an operator, say $O$,
%\begin{align}
% O = \sum_{i,j}O_{ij}\ket{i}\bra{j}.
%\end{align}
%Its state representation is obtained by attaching the extra Hilbert space $\calH_B$ and applying $O_A\times I_B$ on a EPR state on $\calH_A\otimes\calH_B$,
%\begin{align}
%\begin{split}
% (O_A\otimes I_B)\ket{\text{EPR}_{AB}}
% &=(O_A\otimes I_B)\sum_k\frac{\ket{k_A}\otimes\ket{k_B}}{\sqrt{d}}
% = \frac{1}{\sqrt{d}}\sum_{i,j,k}O_{ij}\ket{i_A}\langle j_A|k_A\rangle\otimes\ket{k_B}
% \\
% &=\frac{1}{\sqrt{d}}\sum_{i,j}O_{ij}\ket{i_A}\otimes\ket{j_B}.
%\end{split}
%\end{align}
%On the other hand if we apply $I_A\times (O_B)^t$, then we find
%\begin{align}
%\begin{split}
% (I_A\otimes (O_B)^t)\ket{\text{EPR}_{AB}}
% &=(I_A\otimes O_B)\sum_k\frac{\ket{k_A}\otimes\ket{k_B}}{\sqrt{d}}
% = \frac{1}{\sqrt{d}}\sum_{i,j,k}O_{ji}\ket{k_A}\otimes\ket{i_B}\langle j_B|k_B\rangle
% \\
% &=\frac{1}{\sqrt{d}}\sum_{i,j}O_{ji}\ket{j_A}\otimes\ket{i_B}
% =(O_A\otimes I_B)\ket{\text{EPR}_{AB}}.
%\end{split}
%\end{align}
%This relation is conveniently represented as follows:
%\begin{align}
%\figbox{0.5}{fig_opState} \qquad
%\end{align}
%%
%Now, \eqref{eq:HPstate} may be understood as the state representation of $U$.

Decoding Alice's state is accomplished by applying some operation $V$ to distill the state $R'$ from the collected radiation $D$ and $B'$:
\begin{align}
\figbox{0.4}{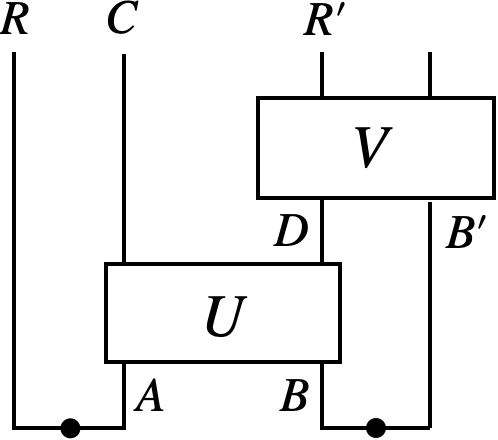} \nonumber
\end{align}
If the state $B'E$ shares nearly maximal amount of information with $R$, then Bob is capable of successfully decoding Alice's state by the quantum recovery channel $V$.

%--------------------------------%
\subsection{Decoding protocol}

\begin{figure}[t]
\centering
\includegraphics[scale=0.7]{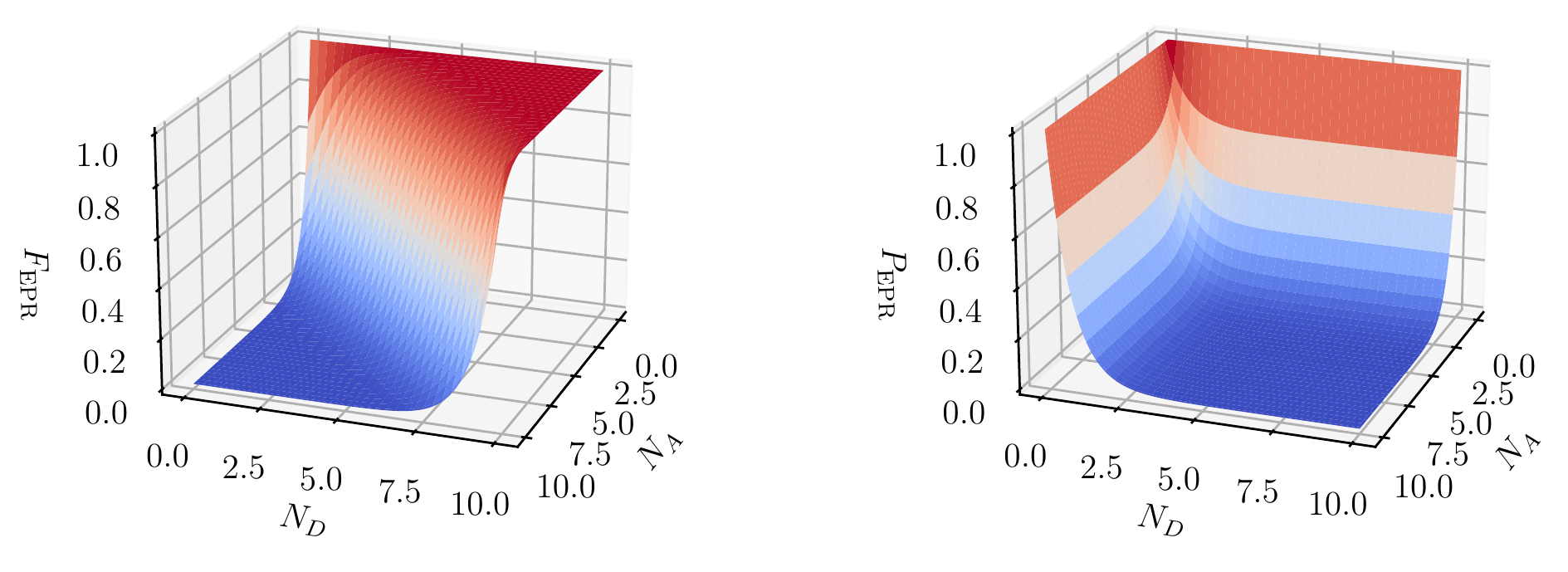}
\caption{The projection probability $P_\text{EPR}$ and decoding fidelity $F_\text{EPR}$ in the ideal case. The total system size is $N=10$, and the subsystem sizes are given by $N_A$ and $N_D$. The left panel shows that the fidelity is close to unity for $N_A\lesssim N_D$.}
\label{fig:idealFP}
\end{figure}

We review how to construct a decoding operation $V$ in the absence of error and noise. Much of this is review of work first performed by~\cite{Yoshida:2017non}.
We assume that Bob has collected all of the early radiation and the late radiation, and has complete knownledge of the unitary driving the black hole's dynamics $U$ (e.g., that he is capable of implementing $U$ on his own qubits for the purpose of decoding).
Given the state $\ket{\Psi_\text{HP}}$~\eqref{eq:HPstate}, Bob's decoding strategy is the following:
\begin{enumerate}
\item Prepare a copy of $\ket{\text{EPR}}_{RA}$, denoted by $\ket{\text{EPR}}_{R'A'}$.

\item Apply $U^\ast$ on $B'A'$. We call this state $\ket{\Psi_\text{in}}$ and it is given by,
\begin{align}
 \ket{\Psi_\text{in}} &= (I_{R}\otimes U_{AB}\otimes U^\ast_{B'A'}\otimes I_{R'}) 
 \ket{\text{EPR}}_{RA}\otimes \ket{\text{EPR}}_{BB'}\otimes \ket{\text{EPR}}_{R'A'},  
 \\ 
 &\rho_\text{in} = \ket{\Psi_\text{in}} \bra{\Psi_\text{in}}=
\figbox{0.4}{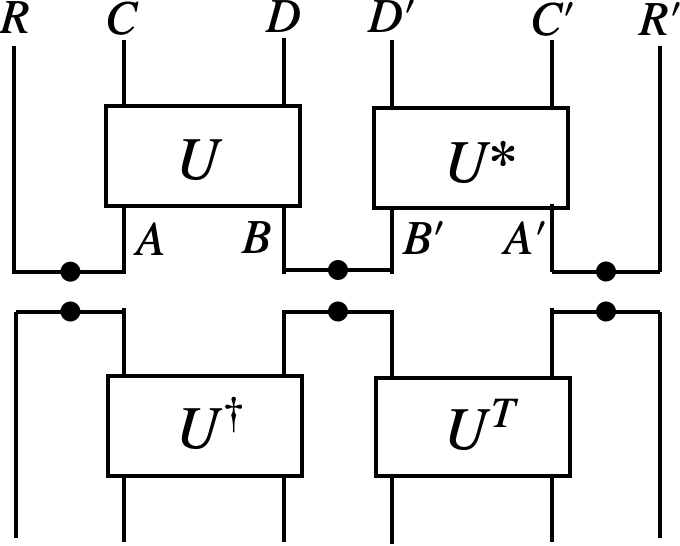} 
\end{align}

\item Project the state onto $\ket{\text{EPR}}_{DD'}$. In other words, Bob repeatedly performs projective measurements on $DD'$ until his state is successfully projected onto $\ket{\text{EPR}}_{DD'}$\footnote{This protocol is called the probabilistic decoding in~\cite{Yoshida:2017non}}.
Letting $\Pi_{DD'}:=\ket{\text{EPR}}\bra{\text{EPR}}_{DD'}$ be the projection operator, we find the probability of getting $\ket{\text{EPR}}_{DD'}$ to be given by
\begin{align}
\label{eq:Pepr}
 P_\text{EPR}=\Tr[\Pi_{DD'}\rho_\text{in}]
 =\frac{1}{d_A^2d_Bd_D}\figbox{0.4}{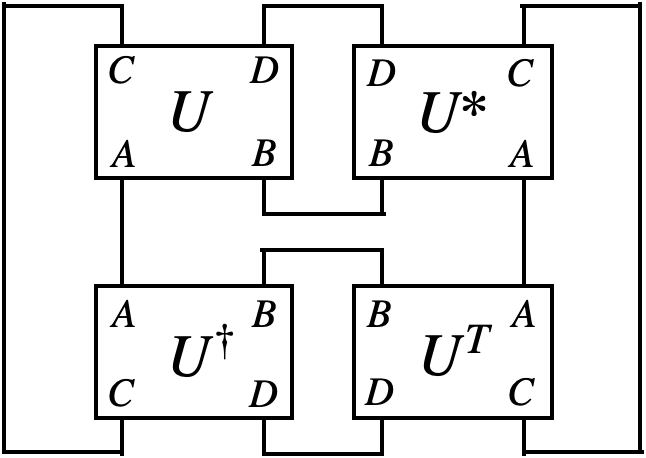} 
\end{align}
The state that is projected onto $\ket{\text{EPR}}_{DD'}$ is
\begin{align}
 \rho_\text{out} = \frac{\Pi_{DD'} \rho_\text{in}\Pi_{DD'}}{P_\text{EPR}} =\frac{1}{P_\text{EPR}}\figbox{0.4}{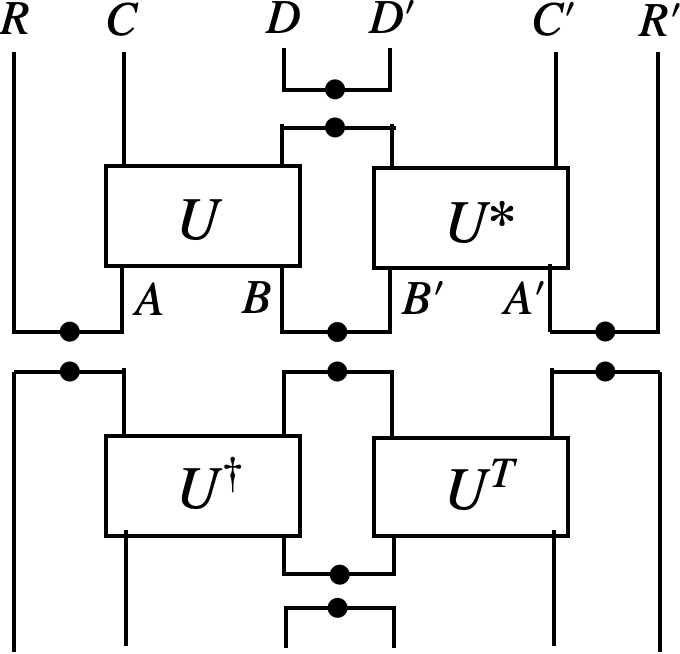} 
\end{align}

For later convenience, we calculate $P_\text{EPR}$ assuming that $U$ is a Haar random unitary operator \footnote{This assumption is justified by the fast scrambling conjecture~\cite{Hayden:2007cs,Sekino:2008he,Shenker:2013pqa,Maldacena:2015waa}. It is technically more correct to take $U$ to be a 2-design as a true Haar random unitary operator takes an exponential number of 2 qubit gates to prepare, but for the sake of quantities that depend only on the second moment such as entanglement entropy this assumption is sufficient~\cite{webb2015clifford,Roberts:2016hpo}. In what follows we will use the Haar random approximation, as it is appropriate for the situation being considered.}:
\begin{align}
\label{eq:PeprIdeal}
 \int\diff U\,P_\text{EPR}=\frac{1}{d^2-1}\left(d_B^2+d_C^2-\frac{d_C^2}{d_A^2}-1\right)
 = \frac{1}{d_A^2} + \frac{1}{d_D^2} -\frac{1}{d_A^2d_D^2} + \calO\left(\frac{1}{d^2}\right).
\end{align}
Here, $\int\diff U$ stands for the integration of the unitary operator $U$ over the Haar measure (see \eqref{eq:barPepr_app} for details of computation).

\item The fidelity between $\rho_\text{out}$ and $\ket{\text{EPR}}_{RR'}$ quantifies the quality of decoding (see Sec.~\ref{sec:ideal_mutual}). It is computed as,
\begin{align}
\label{eq:fid_ideal}
\begin{split}
 F_\text{EPR}&=\Tr[\Pi_{RR'}\rho_\text{out}]
 \\
 &=\frac{1}{P_\text{EPR}d_A^3d_Bd_D}\figbox{0.4}{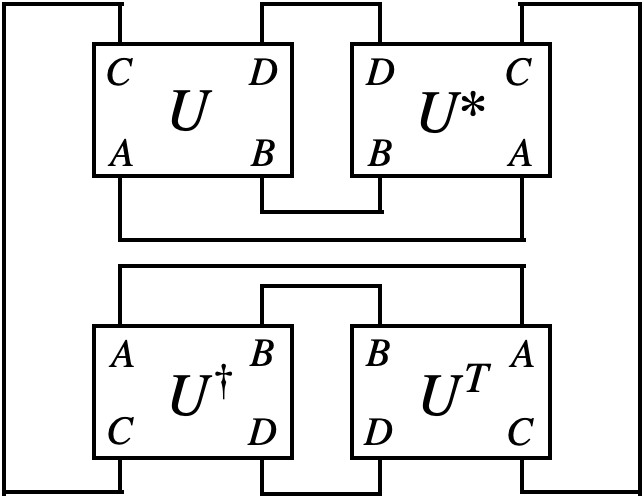} 
 = \frac{d_Cd_D}{P_\text{EPR}d_A^3d_B}
 = \frac{1}{P_\text{EPR}d_A^2}.
\end{split}
\end{align}
Thus, small $P_\text{EPR}$ leads to high fidelity.
In particular, provided $U$ is the Haar random unitary and $d_A\ll d_D$ is satisfied, \eqref{eq:PeprIdeal} reduces to $P_\text{EPR}\approx 1/d_A^{2}$, implying nearly maximal decoding quality, $F_\text{EPR}\approx1$ (see Fig.~\ref{fig:idealFP}).

\end{enumerate}

%--------------------------------%
\subsection{Decoding fidelity and mutual information}
\label{sec:ideal_mutual}

The decoding quality is quantified by mutual information between the reference system $R$ and the collected radiation $B'D$~\eqref{eq:HPstate}.
Here, we make a connection between the mutual information and decoding fidelity~\eqref{eq:fid_ideal}. For computational convenience we consider R\'{e}nyi-2 mutual information,
\begin{align}
 I^{(2)}(R,B'D) := S^{(2)}(R) + S^{(2)}(B'D) - S^{(2)}(RB'D),
\end{align}
with the R\'{e}nyi-2 entropy $S^{(2)}:=-\log\Tr[\rho^2]$.
We find $S^{(2)}(R)=\log d_R=\log d_A$ because the state $R$ is maximally entangled with $A$. The base of logarithm is taken to be 2 throughout the article. Also, $S^{(2)}(RB'D)=\log d_C$ because $RB'CD$ is a pure state and $RB'D$ is maximally entangled with $C$.
Let us compute $\Tr[\rho_{B'D}^2]=\e^{-S^{(2)}(B'D)}$. The state $\rho_{B'D}$ is given by,
\begin{align}
\rho_{B'D} = \Tr_{RC}[\rho_{RB'CD}]
=\frac{1}{d_A}\figbox{0.4}{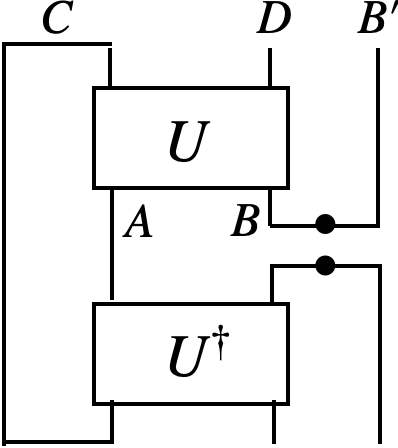}
\end{align}
Then, $\Tr[\rho_{B'D}^2]$ is related to $P_\text{EPR}$~\eqref{eq:Pepr},
\begin{align}
\Tr[\rho_{B'D}^2]
=\frac{1}{d_A^2d_B^2}\figbox{0.3}{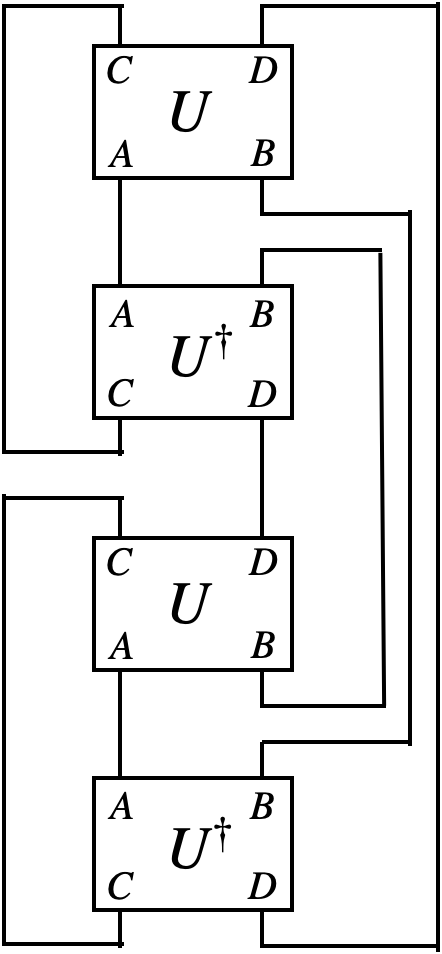} 
=\frac{1}{d_A^2d_B^2}\figbox{0.4}{fig_Pepr} 
=\frac{d_D}{d_B}P_\text{EPR}.
\end{align}
Using $d=d_Ad_B=d_Cd_D$ we find
\begin{align}
\label{eq:PeprMutualIdeal}
P_\text{EPR} 
= \frac{d_C}{d_A}\Tr[\rho_{B'D}^2]
= e^{S^{(2)}(RB'D)-S^{(2)}(R)-S^{(2)}(B'D)}
=e^{-I^{(2)}(R,B'D)},
\end{align}
namely, the projection probability $P_\text{EPR}$ directly measures the R\'enyi-2 mutual information.
We remark that $P_\text{EPR}$ is equal to the ``averaged out-of-time-ordered correlator (OTOC)'', and thus, gives another perspective of information scrambling (see~\cite{Yoshida:2018vly} for details).
Therefore, the decoding fidelity is expressed in terms of $I^{(2)}(R,B'D)$,
\begin{align}
 F_\text{EPR} = \frac{e^{I^{(2)}(R,B'D)}}{d_A^2},
\end{align}
which shows that large $I^{(2)}(R,B'D)$ results in high decoding fidelity~\cite{Yoshida:2018vly} as originally argued based on the decoupling principle~\cite{Hayden:2007cs}.

%%%%%%%%%%%%%%%%%%%%%%%%%%%%%%%%%%%%%%%%%%%%%%%%%
\section{Hayden-Preskill decoding with erasure errors}

In real world physical systems, however, the decoding protocol can be subject to various errors.
In the present work, we are interested in storage errors that occur while Bob keeps early radiation on $B'$ until he applies the decoding protocol upon gathering the late radiation.
\begin{align}
 \figbox{0.4}{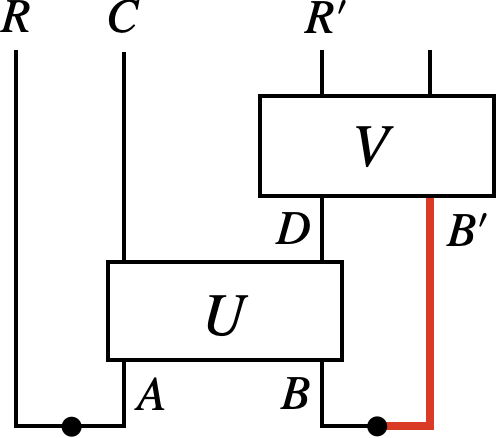}
\end{align}
In this graphical representation, $V$ is the decoding operation that Bob performs on the collected radiation on $B'D$ to distill the EPR pair on $RR'$, and the red wire represents the noisy storage.

%-------------------%
\subsection{Decoding protocol with erasure errors}

We model the noise in storage by erasure errors, {\it i.e.} the situation where some qubits in the storage on $B'$ becomes inaccessible because they are lost or damaged. We will examine the effect of decoherence in Sec.~\ref{sec:erasure}.
The erasure is modeled by randomly choosing qubits to be discarded with probability $p\ (0\le p\le 1)$.
The dimension of the lost Hilbert space on $B'_2$ is $d_{B_2}=2^{N_{B_2}}=2^{pN_B}=(d_B)^p$. The entire state before the application of recovery protocol takes the form,
\begin{align}
 \label{eq:HPstateErase}
 \figbox{0.4}{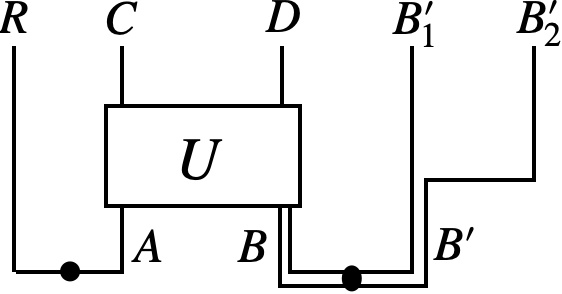}
\end{align}
We graphically doubled the wire between $B$ and $B'$ in order to separate them into $N_{B_2}$-erased qubits on $B'_2$ from $N_{B_1}$-surviving qubits on $B'_1$. 
The state on $B_2'$ is traced out as Bob is unable to access the lost qubits.

Following the protocol in the previous section, Bob applies $U^\ast$ after attaching the EPR state $\ket{\text{EPR}}_{A'R'}$ and the $N_{B_2}$-qubit maximally mixed state, $I_{B_2}/d_{B_2}$, to fill in the erased qubits.
Thus, the associated density operator $\rho_\text{in}$ is graphically represented by,
\begin{align}
\label{eq:rhoInErase}
 \rho_\text{in} = \frac{1}{d_{B_2}^2}\figbox{0.4}{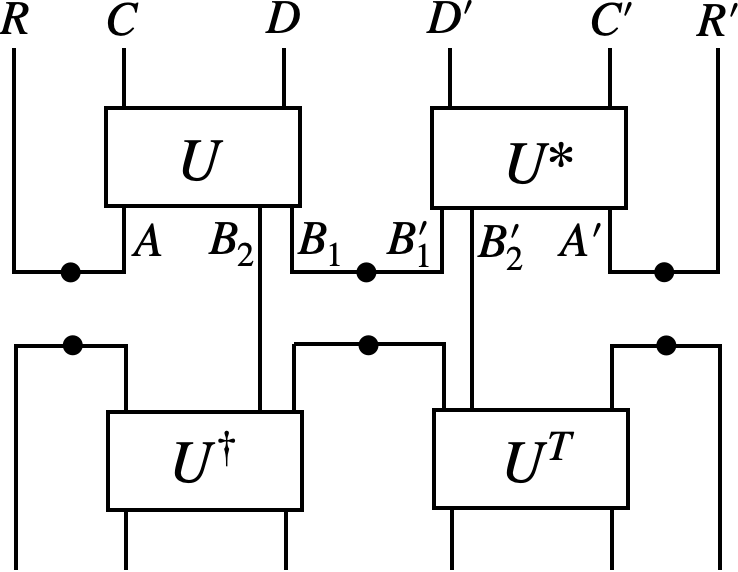}
\end{align}
Then, Bob projects the state~\eqref{eq:rhoInErase} onto $\ket{\text{EPR}}_{DD'}$ to obtain,
\begin{align}
    \rho_\text{out} = \frac{\Pi_{DD'}\rho_\text{in}\Pi_{DD'}}{P_\text{EPR}},
    \qquad
    P_\text{EPR}=\Tr[\Pi_\text{DD'}\rho_\text{in}]
    =\frac{1}{d_A^2d_{B_1}d_{B_2}^2d_D}\figbox{0.4}{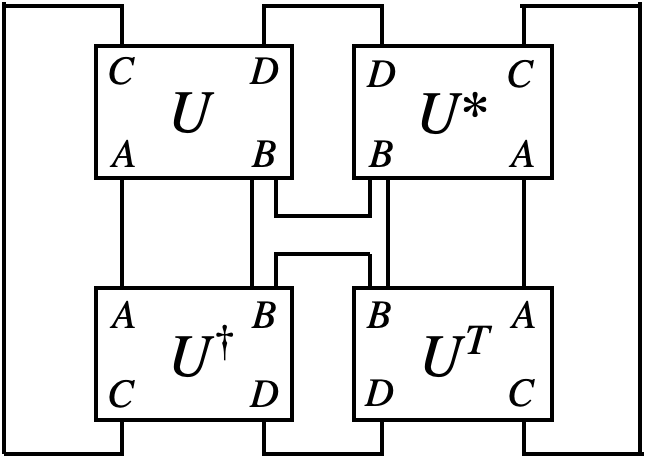}
\end{align}
Upon projecting $\rho_\text{out}$ onto $\ket{\text{EPR}}_{RR'}$ we find the decoding fidelity $F_\text{EPR}=\Tr[\Pi_{RR'}\rho_\text{out}]$ to be
\begin{align}
\label{eq:FeprErase}
 &F_\text{EPR}P_\text{EPR} =\frac{\delta}{d_A^2},
 \qquad
 \delta:=\frac{1}{d_Ad_{B_1}d_{B_2}^2d_D}\figbox{0.4}{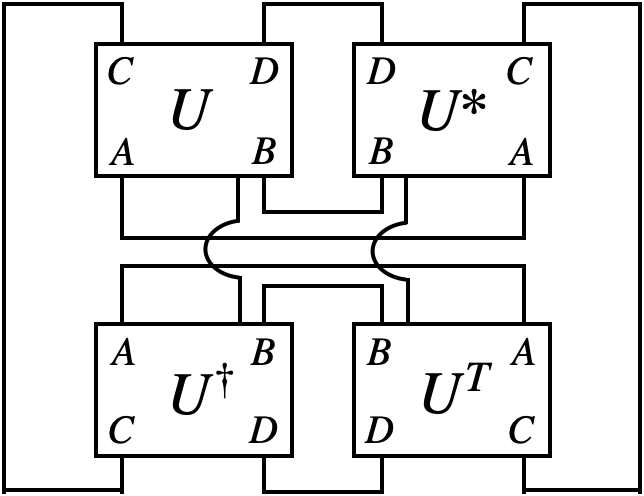}
\end{align}
Successfully projected state is the distilled EPR pair on $RR'$.
Compared with the fidelity in the ideal case~\eqref{eq:fid_ideal}, we note that $\delta$ quantifies the effect of error in $F_\text{EPR}P_\text{EPR}$.

%-------------------%
\subsection{Mutual information}

In order to make a connection to the R\'enyi-2 mutual information $I^{(2)}(R,B'_1D)=S^{(2)}(R)+S^{(2)}(B'_1D)-S^{(2)}(RB'_1D)$, we firstly compute $\Tr[(\rho_{RB'_1D})^2]=2^{-S^{(2)}(RB'_1D)}$ on the state~\eqref{eq:HPstateErase},
\begin{align}
\label{eq:rhoRBD2erase}
 \Tr[(\rho_{RB'_1D})^2] = \frac{1}{d_A^2d_B^2}\figbox{0.3}{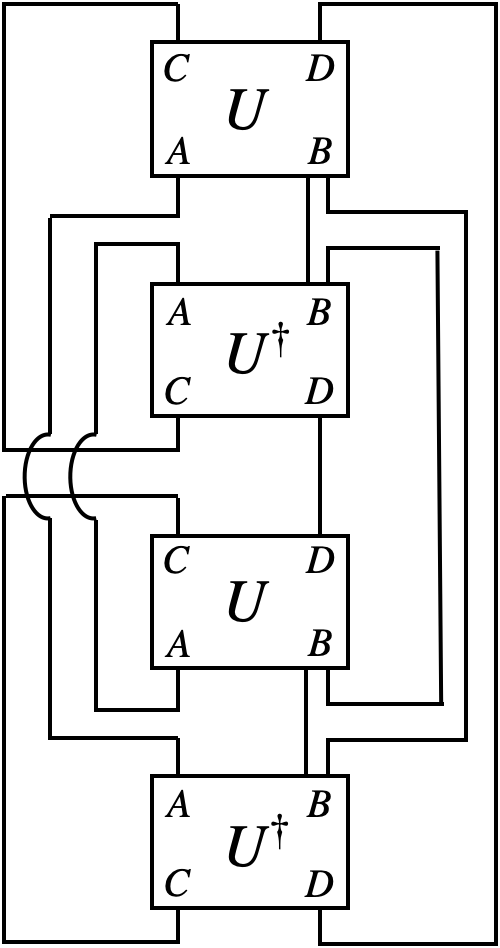}
 =\frac{d_{B_2}}{d_C}\delta.
\end{align}
This equation is recast into the following form,
\begin{align}
\label{eq:deltaMutualErase}
 \delta = 2^{-S^{(2)}(RB'_1D)+S^{(2)}(C)-S^{(2)}(B'_2)}
 =\frac{2^{S^{(2)}(R)+S^{(2)}(B'_1D)-S^{(2)}(RB'_1D)}}
 {2^{S^{(2)}(R)+S^{(2)}(B'_1D)-S^{(2)}(C)+S^{(2)}(B'_2)}}
 =\frac{2^{I^{(2)}(R,B'_1D)}}
 {2^{S^{(2)}(D)+S^{(2)}(B'_1D)-S^{(2)}(B'_1)}},
\end{align}
where we have used $S^{(2)}(R)+S^{(2)}(B'_1)+S^{(2)}(B'_2)=S^{(2)}(C)+S^{(2)}(D)$.

We next compute $\Tr[(\rho_{B'_1D})^2]=2^{-S^{(2)}(B'_1D)}$,
\begin{align}
\label{eq:rhoBD2erase}
 \Tr[(\rho_{B'_1D})^2]=\frac{1}{d_A^2d_B^2}\figbox{0.3}{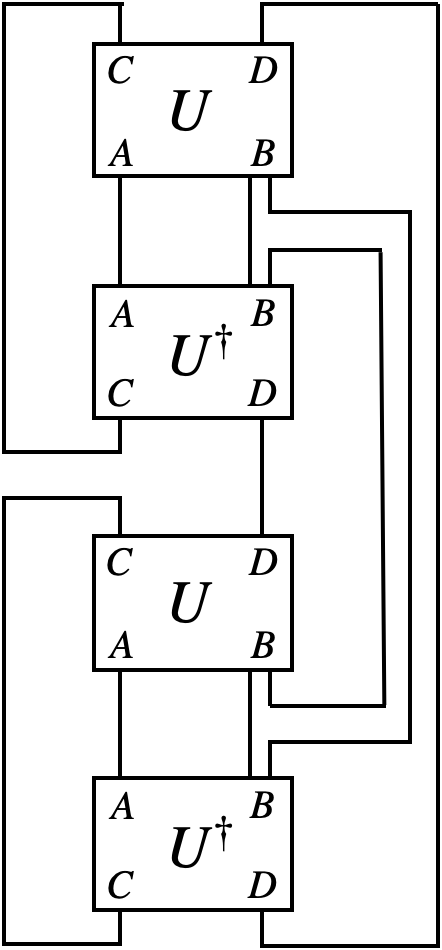}
 =\frac{d_D}{d_{B_1}}P_\text{EPR}.
\end{align}
This can be written as
\begin{align}
\label{eq:Pepr_I_Erase}
 P_\text{EPR} = 2^{-S^{(2)}(D)-S^{(2)}(B'_1D)+S^{(2)}(B'_1)}.
\end{align}
With the erasure error, $P_\text{EPR}$ is given by a combination of R\'enyi-2 entropies instead of R\'enyi-2 mutual information, the latter of which encodes the recoverability of Alice's state. The ratio of these quantities is precisely captured by the error factor $\delta$~\eqref{eq:deltaMutualErase}~\cite{Yoshida:2018vly}.

Using \eqref{eq:FeprErase}, \eqref{eq:deltaMutualErase}, and \eqref{eq:Pepr_I_Erase} we find
\begin{align}
\label{eq:Fepr_I_Erase}
F_\text{EPR} = \frac{2^{I^{(2)}(R,B'_1D)}}{d_A^2}.
\end{align}
Even though $P_\text{EPR}$ takes a different form from the ideal case~\eqref{eq:PeprMutualIdeal}, the decoding fidelity rightly reflects the mutual information between the state on $R$ and the collected radiation $B'_1D$, to which Bob has an access. The effect of erasure is translated to the reduction of the mutual information and vice versa.
%-------------------%

\subsection{Decoding fidelity}

Let us compute $\delta$~\eqref{eq:FeprErase} provided that the time evolution is governed by a Haar random unitary,
\begin{align}
\begin{split}
 \delta =
 \frac{1}{d_Ad_{B_1}d_{B_2}^2d_D}U_{a_1(b_1b_2)c_1d_1}U^\ast_{a_1(b_1b'_2)c_2d_1}U_{a_2(b'_1b'_2)c_2d_2}U^\ast_{a_2(b'_1b_2)c_1d_2}.
\end{split}
\end{align}
Executing the Haar integral~\eqref{eq:barDeltaErase_app} we find the Haar random average of $\delta$,
\begin{align}
\label{eq:barDeltaErase}
\bar{\delta}:= \int\diff U\,\delta =\frac{1}{d^2-1}\left[\frac{d^2}{d_B^{2p}} + d_C^2
 -\frac{d_C^2}{d_B^{2p}}-1\right]
 = \frac{1}{d_B^{2p}} + \frac{1}{d_D^2}\left(1-\frac{1}{d_B^{2p}}\right) + \calO\left(\frac{1}{d^2}\right).
\end{align}
It is reduced to 1 without erasure errors, $p=0$.
For small $p$, using $d_B^{-2p}=1-p(2\ln 2 \log d_B) + \calO(p^2)$ we get,
\begin{align}
\label{eq:barDeltaErase_approx}
\bar{\delta} \approx 1-p(2\ln 2 \log d_B)\left(1-\frac{1}{d_D^2}\right).
\end{align}
The error factor is approximately proportional to the number of erased qubits $N_{B_2}=p\log d_B$.

$P_\text{EPR}$ is similarly computed:
\begin{align}
\begin{split}
 P_\text{EPR} =
 \frac{1}{d_A^2d_{B_1}d_{B_2}^2d_D}U_{a_1(b_1b_2)c_1d_1}U^\ast_{a_2(b_1b'_2)c_2d_1}U_{a_2(b'_1b'_2)c_2d_2}U^\ast_{a_1(b'_1b_2)c_1d_2}.
\end{split}
\end{align}
Its Haar average~\eqref{eq:barPeprErase_app} results in,
\begin{align}
\label{eq:barPeprErase}
\bar{P}_\text{EPR}:= \int\diff U\,\bar{P}_\text{EPR}
=\frac{1}{d^2-1}
\left[d_B^{2(1-p)} + d_C^2-\frac{d_C^2}{d_A^2d_B^{2p}}-1\right]
 = \frac{1}{d_A^2d_B^{2p}} + \frac{1}{d_D^2} -\frac{1}{d_A^2d_B^{2p}d_D^2} + \calO\left(\frac{1}{d^2}\right).
\end{align}
Therefore, the decoding fidelity is
\begin{align}
\label{eq:FeprEraseHaar}
F_\text{EPR} = \frac{\bar{\delta}}{d_A^2\bar{P}_\text{EPR}}
= \frac{d_D^2+d_B^{2p}-1}{d_D^2+d_A^2d_B^{2p}-1} + \calO\left(\frac{1}{d^2}\right).
\end{align}

%%%%%%%%%%%%%%%%%%%%%%%%%%%%%%%%%%%%%%%%%%%%%%%%%
\section{Hayden-Preskill decoding with decoherence in storage}
\label{sec:erasure}

We consider the effects of two types of errors: decoherence in storage of the collected early radiation on $B'$, and imperfect implementation of the backward unitary evolution combined with decoherence.

%--------------------------------%
\subsection{Decoherence in storage}

We model the decoherence by the depolarizing channel,
\begin{align}
 \calQ(\rho) = (1-p)\rho + p\frac{I}{d}\Tr\rho.
\end{align}
Although the decoherence affects the state on $B'$, the quantum channel may be understood to act on $\ket{\text{EPR}}\bra{\text{EPR}}_{BB'}$,\footnote{This identity is checked as follows:
\begin{align}
\begin{split}
 \calQ_{B'}(\ket{\text{EPR}}\bra{\text{EPR}}_{BB'})
 &= \frac{1}{d_B}\sum_{i,j}\ket{i_B}\bra{j_B}\otimes\calQ(\ket{i_{B'}}\bra{j_{B'}})
 \\
 &= (1-p)\ket{\text{EPR}}\bra{\text{EPR}}_{BB'}
 +\frac{p}{d_B^2}\sum_{i,j}\Tr[\ket{i_{B'}}\bra{j_{B'}}](\ket{i_B}\bra{j_B}\otimes I)
 \\
 &= (1-p)\ket{\text{EPR}}\bra{\text{EPR}}_{BB'} + p\frac{I\otimes I}{d_B^2}
 \\
 &=\calQ_{BB'}(\ket{\text{EPR}}\bra{\text{EPR}}_{BB'}).
\end{split}
\end{align}
}
\begin{align}
 &\calQ_{B'}(\ket{\text{EPR}}\bra{\text{EPR}}_{BB'})
 =\calQ_{BB'}(\ket{\text{EPR}}\bra{\text{EPR}}_{BB'}).
\end{align}
The decoherence in storage is responsible for the decay of entanglement between the early radiation $B'$ and the black hole $B$, that are initially maximally entangled, resulting in the state,
\begin{align}
\label{eq:HPstateDecoh}
 \figbox{0.35}{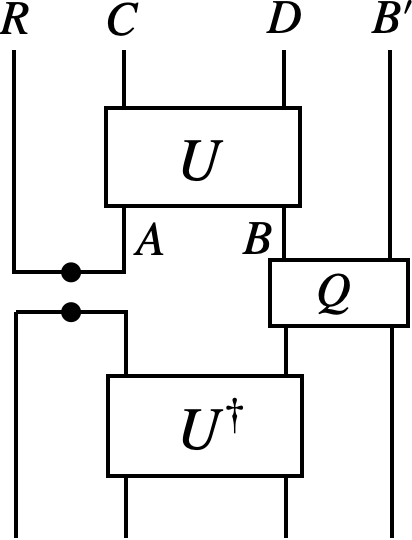}
\end{align}
To start decoding, Bob introduces an EPR pair on $R'A'$ followed by an application of $U^\ast$ to obtain the state,
\begin{align}
 \rho_\text{in} 
 &= (U_{AB}\otimes U_{B'A'}^\ast)\calQ_{BB'}\big(\ket{\text{EPR}}\bra{\text{EPR}}_{RA}\otimes\ket{\text{EPR}}
 \bra{\text{EPR}}_{BB'}\otimes\ket{\text{EPR}}\bra{\text{EPR}}_{R'A'}\big)(U_{AB}^\dag\otimes U_{B'A'}^T)
 \nonumber\\
 &=\figbox{0.4}{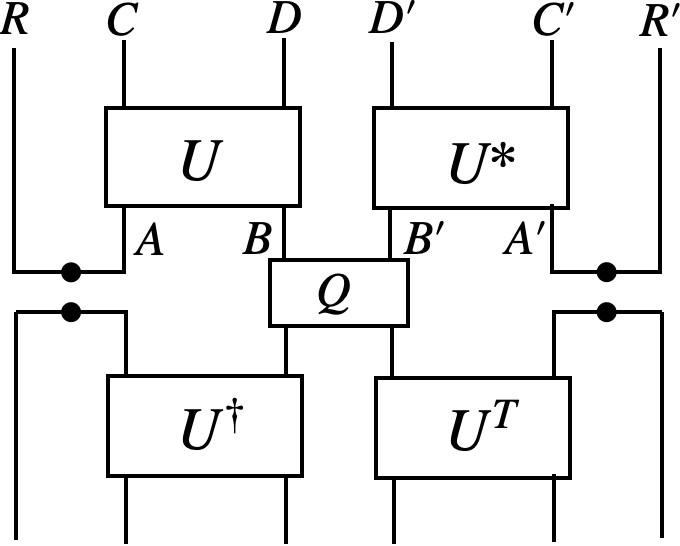}
\end{align}
Sequential projections of $\rho_\text{in}$ on $\ket{\text{EPR}}_{DD'}$ and then on $\ket{\text{EPR}}_{RR'}$ leads to the decoding fidelity,
\begin{align}
\label{eq:FeprDecoh}
 &F_\text{EPR}P_\text{EPR} =\frac{\delta}{d_A^2},
 \qquad
 \delta:=\frac{1}{d_Ad_D}\figbox{0.4}{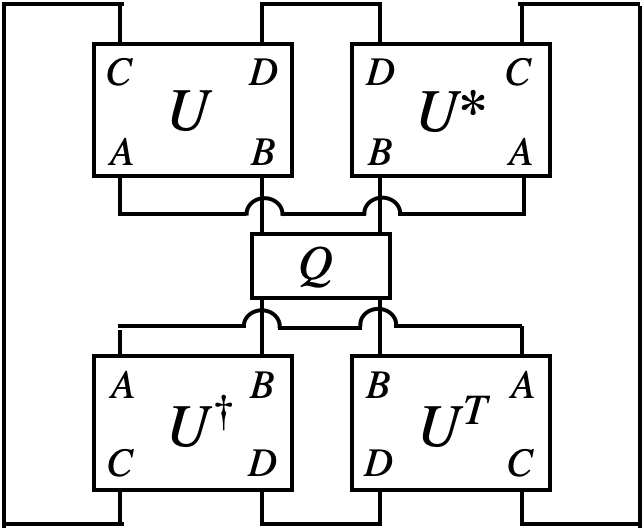},
\end{align}
where $P_\text{EPR}$ is given by,
\begin{align}
 P_\text{EPR} = \frac{1}{d_A^2d_D}\figbox{0.4}{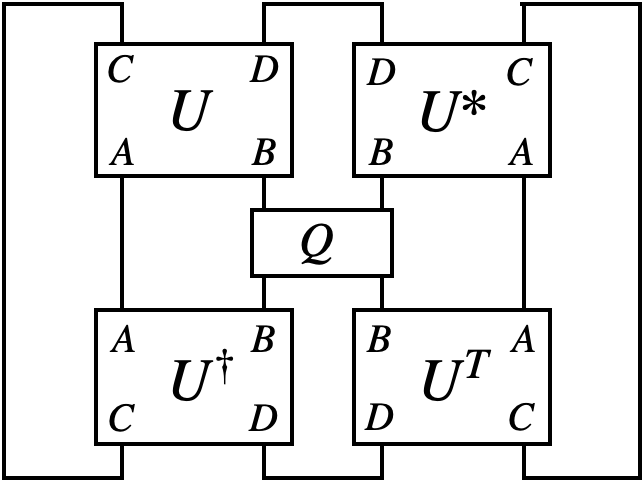}
\end{align}
Below, we will discuss its relation to the mutual information, and calculate these quantity explicitly.

\subsubsection*{Mutual information}

Let us relate the decoding fidelity to the R\'enyi-2 mutual information. $\Tr[(\rho_{RB'D})^2]=2^{-S^{(2)}(RB'D)}$ on the state~\eqref{eq:HPstateDecoh} is converted to,
\begin{align}
\label{eq:rhoRBD2decoh}
 \Tr[(\rho_{RB'D})^2] = \frac{1}{d_A^2}\figbox{0.3}{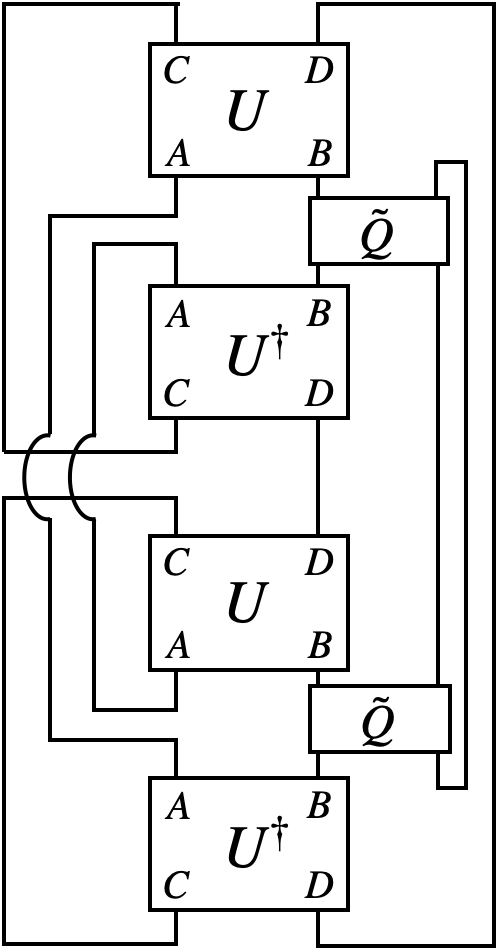}
 = \frac{1}{d_A^2}\figbox{0.35}{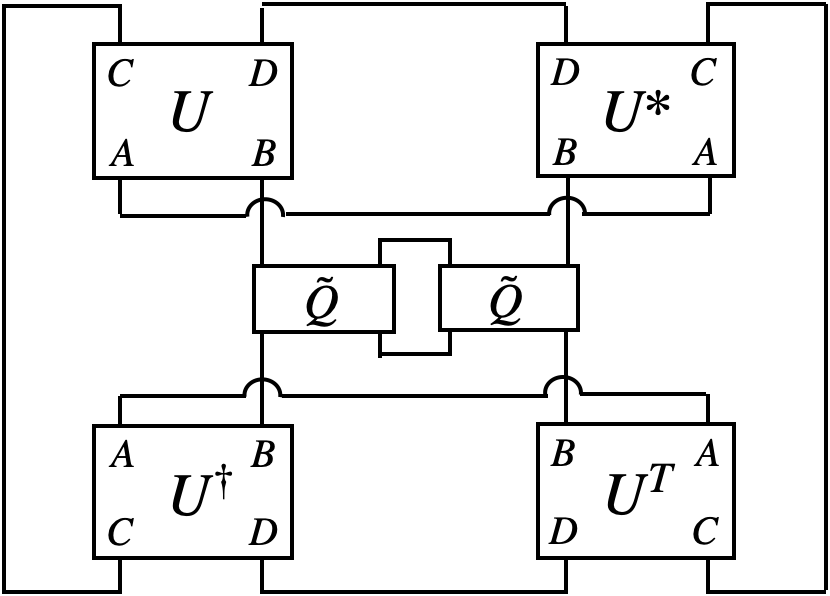}
 =\frac{\delta}{d_C}.
\end{align}
In the last equality we used,
\begin{align}
 &\figbox{0.4}{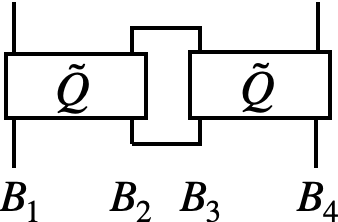}
 \quad=\quad
 \figbox{0.4}{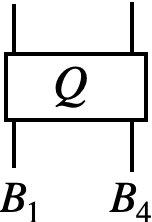}
\end{align}
where $\tilde{Q}$ describes the depolarizing channel with probability $\tilde{p}$ satisfying (see Appendix~\ref{app:relation_decoh} for derivation),
\begin{align}
\label{eq:tildeP}
2\tilde{p}-\tilde{p}^2=p.
\end{align}
The relation~\eqref{eq:rhoRBD2decoh} implies the following,
\begin{align}
 \delta = 2^{-S^{(2)}(RB'D)+S^{(2)}(C)}
 =\frac{2^{S^{(2)}(R)+S^{(2)}(B'D)-S^{(2)}(RB'D)}}
 {2^{S^{(2)}(R)+S^{(2)}(B'D)-S^{(2)}(C)}}
 =\frac{2^{\tilde{I}^{(2)}(R,B'D)}}
 {2^{S^{(2)}(D)+S^{(2)}(B'D)-S^{(2)}(B')}}.
\end{align}
We have used $\tilde{I}^{(2)}(R,B'D)=S^{(2)}(R)+S^{(2)}(B'D)-S^{(2)}(RB'D)$ and $d=2^{S^{(2)}(R)+S^{(2)}(B')}=2^{S^{(2)}(C)+S^{(2)}(D)}$. The tilde on $I^{(2)}$ is to emphasize that the depolarizing channel $\tilde{Q}$ is employed for the computation.

We next compute $\Tr[(\rho_{B'D})^2]=2^{-S^{(2)}(B'D)}$,
\begin{align}
\label{eq:rhoBD2decoh}
 \Tr[(\rho_{B'D})^2]
 =\frac{1}{d_A^2}\figbox{0.3}{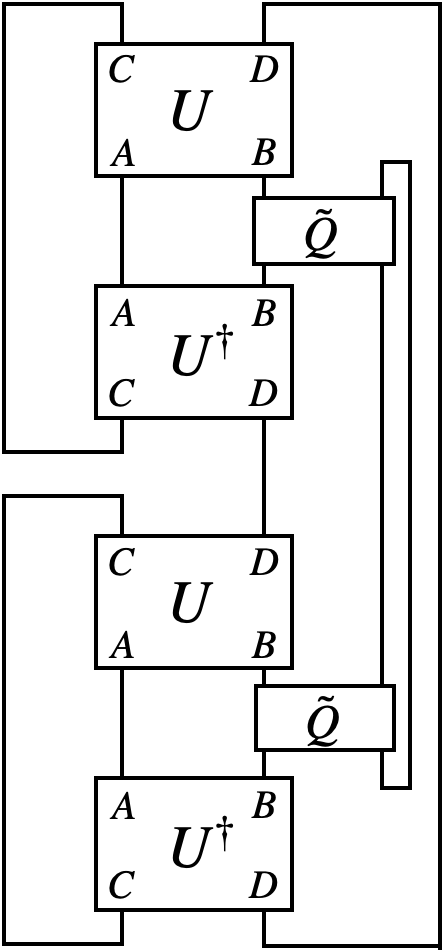}
 =\frac{1}{d_A^2}\figbox{0.35}{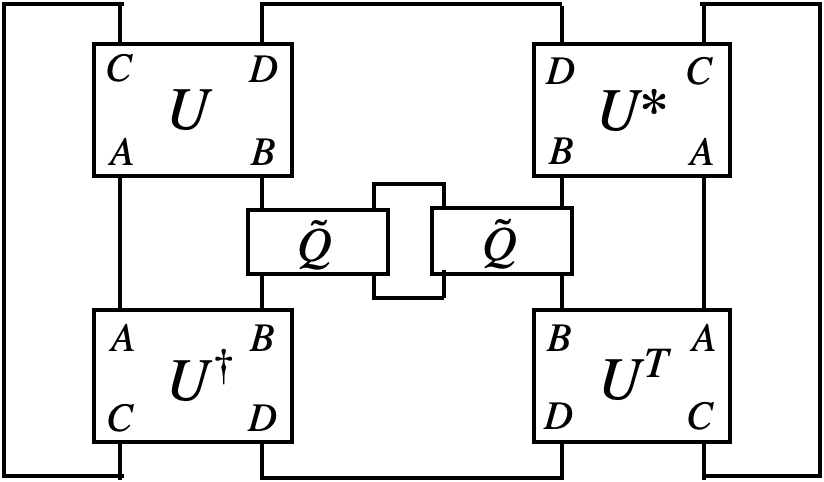}
 =\frac{d_D}{d_B}P_\text{EPR}.
\end{align}
This is equivalently written as
\begin{align}
 P_\text{EPR} = 2^{-S^{(2)}(D)-S^{(2)}(B'D)+S^{(2)}(B')}
\end{align}
which, in turn, implies
\begin{align}
F_\text{EPR} = \frac{2^{\tilde{I}^{(2)}(R,B'D)}}{d_A^2}.
\end{align}
Therefore, the larger the R\'enyi-2 mutual information is the better the decoding quality becomes. Note, however, that the mutual information is computed by using the quantum channel $\tilde{Q}$ with $\tilde{p}\le p$~\eqref{eq:tildeP}. The intuition here is that the $\tilde{Q}$ channel involves participation from $B_2$ and $B_3$ as ancilla qubits, which can in general change the mutual information of states after application of a quantum channel even of systems non involving the ancilla when contrasted with the $Q$ channel which does not involve the ancillas at all. In this sense, \eqref{eq:tildeP} is a consistency condition that must be satisfied to ensure that the ancilla qubits have no affect on the mutual information when the $\tilde{Q}$ channel is applied.

\subsubsection*{Decoding fidelity}

Let us compute $\delta$:
\begin{align}
\begin{split}
 \delta 
 &= \frac{1-p}{d_Ad_Bd_D}\figbox{0.4}{fig_Fepr} + \frac{p}{d_Ad_B^2d_D}\figbox{0.4}{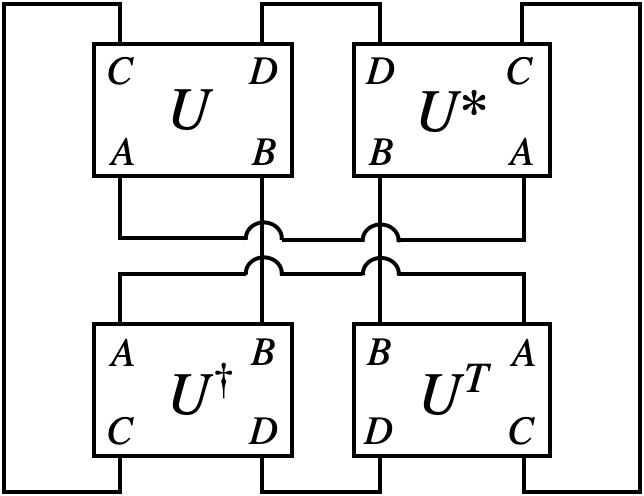} 
 \\
 &= (1-p)
 + \frac{p}{d_Ad_B^2d_D}U_{a_1b_1c_1d_1}U^\ast_{a_1b_2c_2d_1}U_{a_2b_2c_2d_2}U^\ast_{a_2b_1c_1d_2}.
\end{split}
\end{align}
In order to further proceed with the computation, we assume that the time evolution is described by Haar random unitary. 
Haar random average of the second term is computed in \eqref{eq:barDeltaDecoh_app} and we find the error rate,
\begin{align}
\label{eq:barDeltaDecoh}
\bar{\delta}:= \int\diff U \delta
= 1-p + \frac{p}{d^2-1}\left[d_A^2+d_C^2-\frac{d_A^2}{d_D^2}-1\right]
= 1- p\left(1-\frac{1}{d_B^2}-\frac{1}{d_D^2}+\frac{1}{d_B^2d_D^2}\right) + \calO\left(\frac{1}{d^2}\right).
\end{align}

Similarly, we carry out the Haar integral to find Haar average of $P_\text{EPR}$:
\begin{align}
\label{eq:barPeprDecoh}
\begin{split}
 \bar{P}_\text{EPR} 
 &:=\int\diff U \bar{P}_\text{EPR}
 = \frac{1-p}{d^2-1}
\left[d_B^2 + d_C^2-\frac{d_C^2}{d_A^2}-1\right]
 + \frac{p}{d_D^2}
 \\
 &= \left[\frac{1}{d_A^2} + \frac{1}{d_D^2}-\frac{1}{d_A^2d_D^2}\right]
 - \frac{p}{d_A^2}\left[1-\frac{1}{d_D^2}\right] + \calO\left(\frac{1}{d^2}\right).
\end{split}
\end{align}
The decoding fidelity is now computed by plugging these into~\eqref{eq:FeprDecoh}.
Its $p$ dependence in comparison with the one with the erasure errors~\eqref{eq:FeprEraseHaar} is shown in Fig.~\ref{fig:F_pdep} and discussed in Sec.~\ref{sec:discussion}.

%--------------------------------%
\subsection{Imperfect noisy backward evolution}

Suppose Bob is only capable of implementing noisy backward time evolution of $U$ by using a matrix $\tilde{U}^\ast$ that is not precisely the complex conjugate of $U$ because of either imperfect implementation, decoherence, or both.
We model the situation by the following initial state:
\begin{align}
 \rho_\text{in} 
 &= U_{AB}\calQ^T_{A'B'}\big(\ket{\text{EPR}}\bra{\text{EPR}}_{RA}\otimes\ket{\text{EPR}}
 \bra{\text{EPR}}_{BB'}\otimes\ket{\text{EPR}}\bra{\text{EPR}}_{R'A'}\big)U_{AB}^\dag
 \nonumber\\
 &=\figbox{0.4}{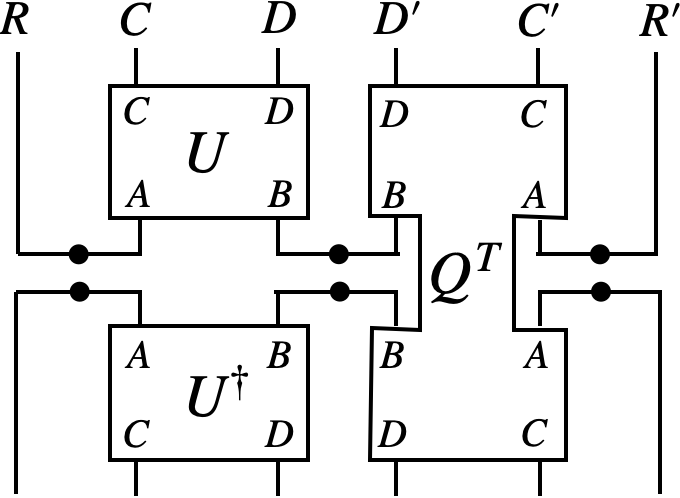}
\end{align}
where the quantum channel used here is
\begin{align}
 \calQ(\rho) = (1-p)\tilde{U}\rho \tilde{U}^\dag + p\frac{I}{d}\Tr\rho.
\end{align}
The decoding fidelity $F_\text{EPR}$ is given by,
\begin{align}
 F_\text{EPR}P_\text{EPR} = \frac{\Delta}{d_A^2},
 \qquad
 \Delta := \frac{1}{d_Ad_Bd_D}\figbox{0.4}{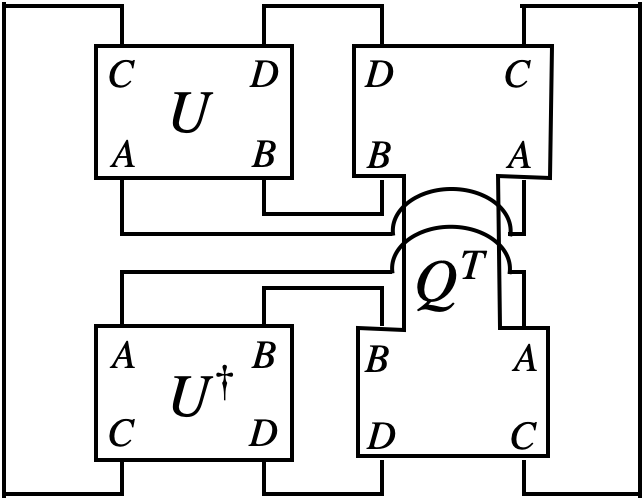},
\end{align}
with the probability of $\rho_\text{in}$ projected onto $\ket{\text{EPR}}_{DD'}$,
\begin{align}
 P_\text{EPR} =\frac{1}{d_A^2d_Bd_D}\figbox{0.4}{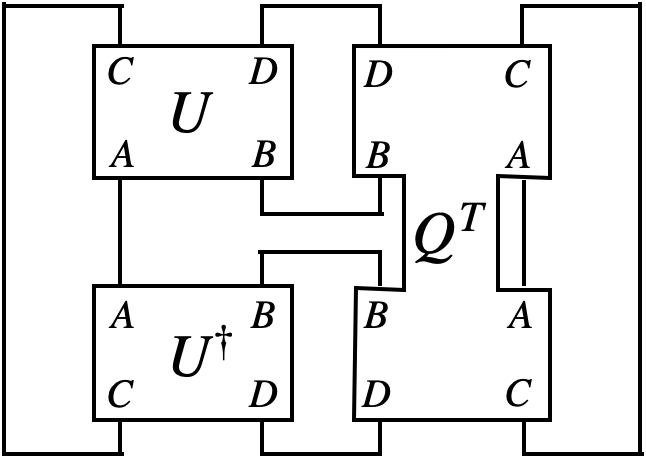}
\end{align}
Then, $\Delta$ is computed as
\begin{align}
\begin{split}
 \Delta 
 &= \frac{1-p}{d_Ad_Bd_D}\figbox{0.4}{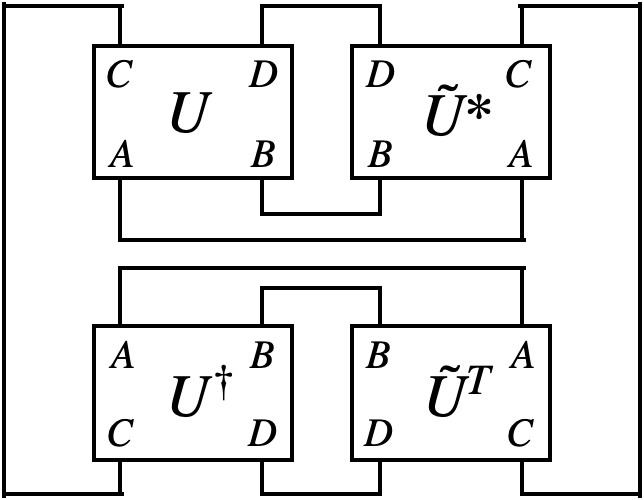} + \frac{p}{d_D^2}
 = (1-p)\eta + \frac{p}{d_D^2},
\end{split}
\end{align}
where
\begin{align}
  \eta = \Tr\left[I\otimes\Pi_{DD'}U\tilde{U}^\dag\left(\frac{I}{d_C}\otimes\Pi_{DD'}\right)\tilde{U}^\dag U\right]
\end{align}
quantifies the difference between $U$ and $\tilde{U}$, and becomes precisely their two-norm overlap when $d_C=1$~\cite{Yoshida:2018vly}.

%%%%%%%%%%%%%%%%%%%%%%%%%%%%%%%%%%%%%%%%%%%%%%%%%
\section{Comparison: Erasure vs Decoherence}
\label{sec:discussion}

\begin{figure}[th]
\centering
\begin{minipage}{.49\textwidth}
\subfloat[$p=0.1$]{
\includegraphics[scale=0.7]{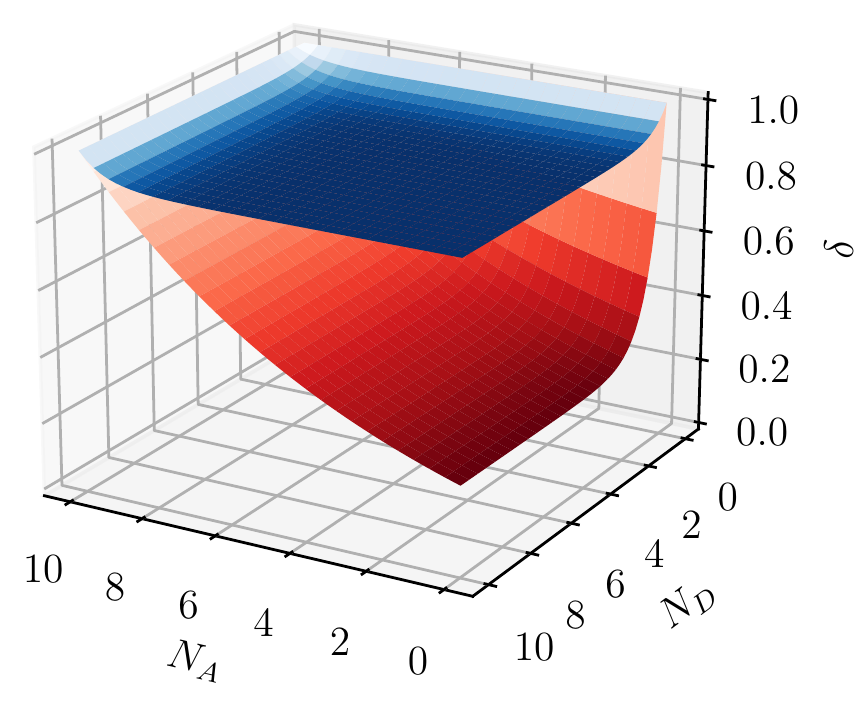}
\label{fig:delta_p01}
}\end{minipage}\
\begin{minipage}{.49\textwidth}
\subfloat[$p=0.2$]{
\includegraphics[scale=0.7]{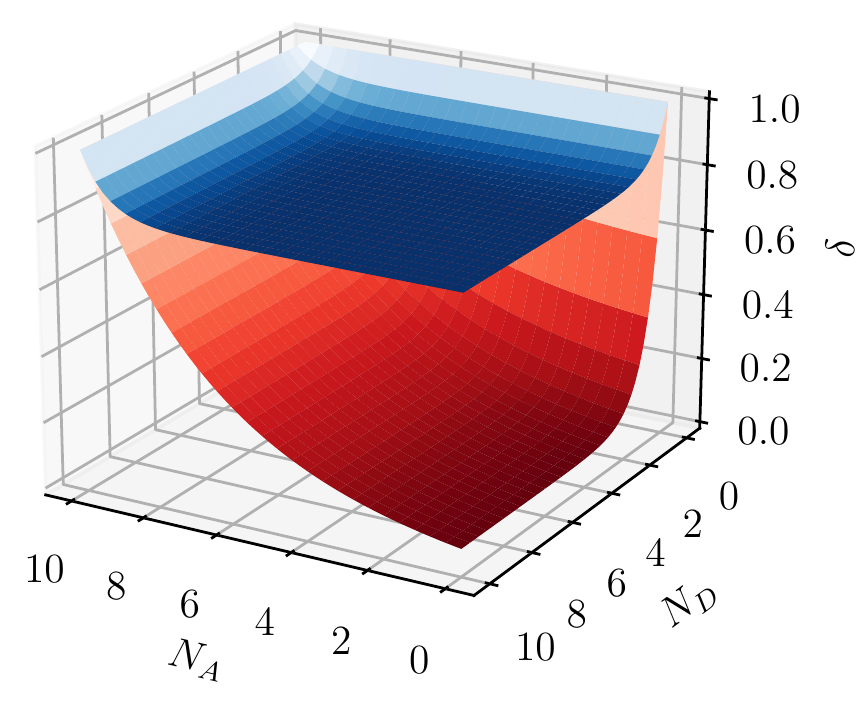}
\label{fig:delta_p02}
}\end{minipage}
\caption{The Haar random average of error factor $\delta$ due to the erasure errors (red surface) and the decoherence (blue surface) in the storage. Without errors the error factor satisfies $\delta=1$. Total system size is $N=10$, and the subsystem sizes are given by $N_A$ and $N_D$.
The erasure error induces bigger depletion in $\delta$ than the decoherence.}
\label{fig:delta}
\end{figure}

As we have discussed, the error factor $\delta$ quantifies the deviation from unity of $F_\text{EPR}P_\text{EPR}$.
Indeed, the error factor is identically unity in the absence of errors.
Comparisons between the erasure errors and the decoherence in terms of the error factor are shown in Fig.~\ref{fig:delta}.
For computations of $\delta$ the Haar random average of unitary evolutions is taken.
Both plots clearly show that the erasure error have severe impacts relative to the decoherence, particularly when the size of Alice's state $N_A$ is small. This is understood from the fact that, for small $p$, the effect of erasure error is proportional to the number of erased qubits $\sim p N_B$, that becomes larger when $N_A$ is smaller~\eqref{eq:barDeltaErase_approx}. For general $p$, the error factor decays exponentially with respect to $p$~\eqref{eq:barDeltaErase} given sufficiently large $N_D$, in contrast to the effect of decoherence~\eqref{eq:barPeprDecoh}, that induces the decay of $\delta$ proportional to $p$.\footnote{Here, we made a comparison between the erasure errors with error probability $p$ and the decoherence with probability $p$. It is, however, not clear whether this comparison with the same $p$ is fair. Nevertheless, we believe our analysis is still useful to see how these effects depend on the parameter $p$.}

\begin{figure}[th]
\centering
\begin{minipage}{.49\textwidth}
\subfloat[Erasure]{
\includegraphics[scale=0.5]{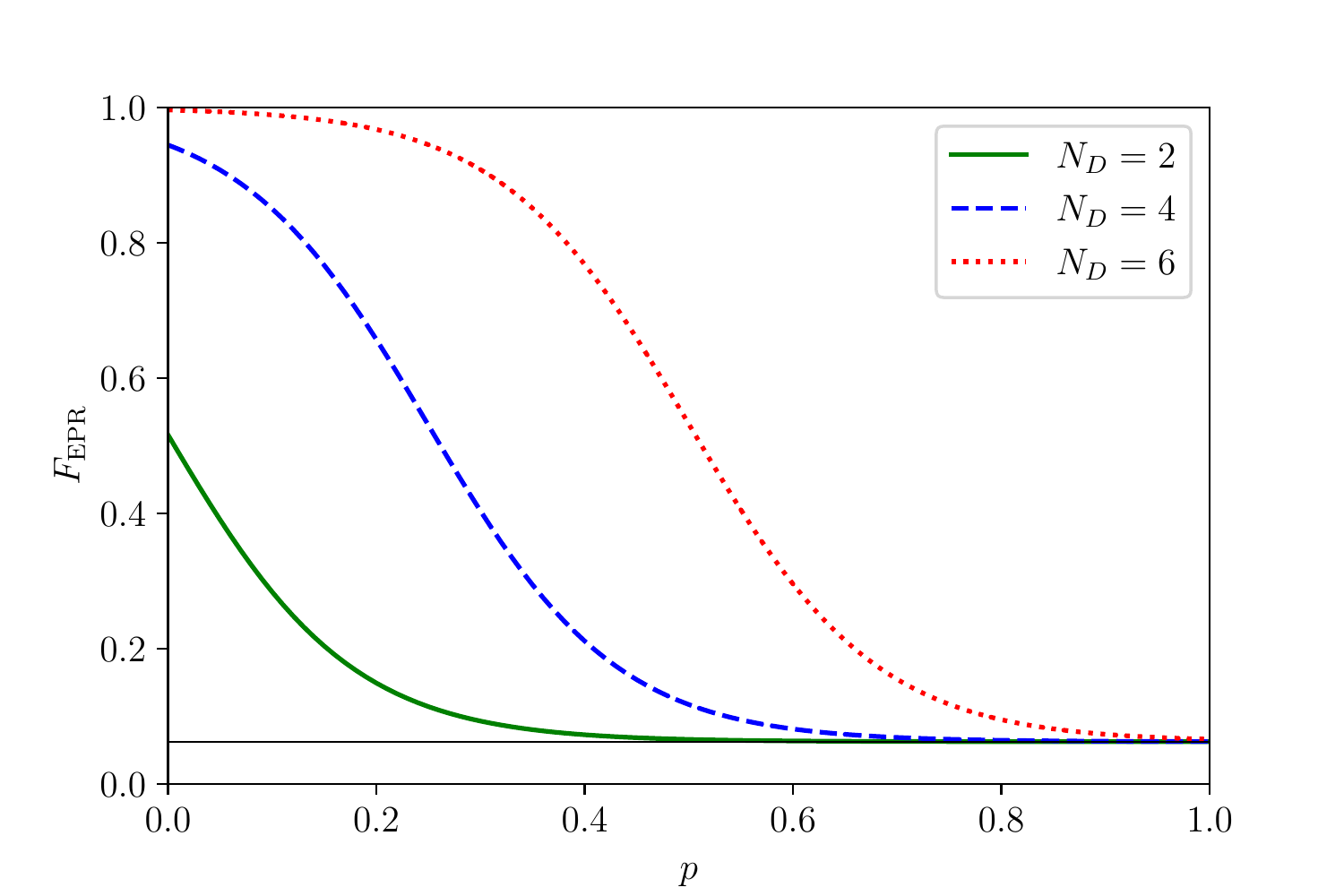}
\label{fig:Ferasure_pdep}
}\end{minipage}\
\begin{minipage}{.49\textwidth}
\subfloat[Decoherence]{
\includegraphics[scale=0.5]{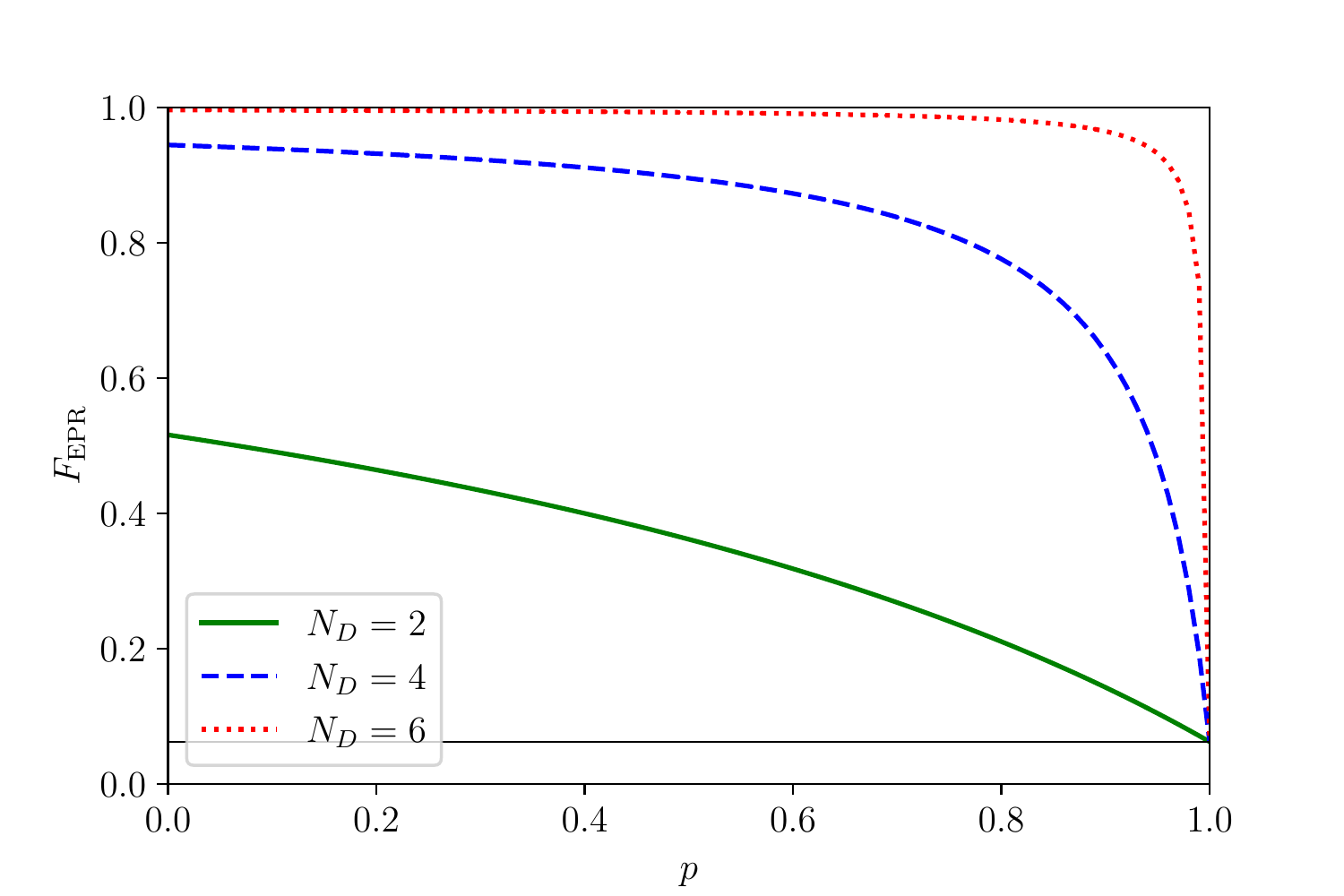}
\label{fig:Fdecoh_pdep}
}\end{minipage}
\caption{Error-probability dependence of the decoding fidelity in the presence of (a) erasure errors and (b) decoherence. The sizes of total system and Alice's state are fixed to be $N=10$ and $N_A=2$, respectively. The black solid horizontal line indicates the lower bound of decoding fidelity, $1/d_A^2$.
The erasure errors severely affect the decoding fidelity compared while the it is well protected against the decoherence effect particularly when $N_D$ is large relative to $N_A$.}
\label{fig:F_pdep}
\end{figure}

\begin{figure}[th]
\centering
\begin{minipage}{.49\textwidth}
\subfloat[Erasure]{
\includegraphics[scale=0.5]{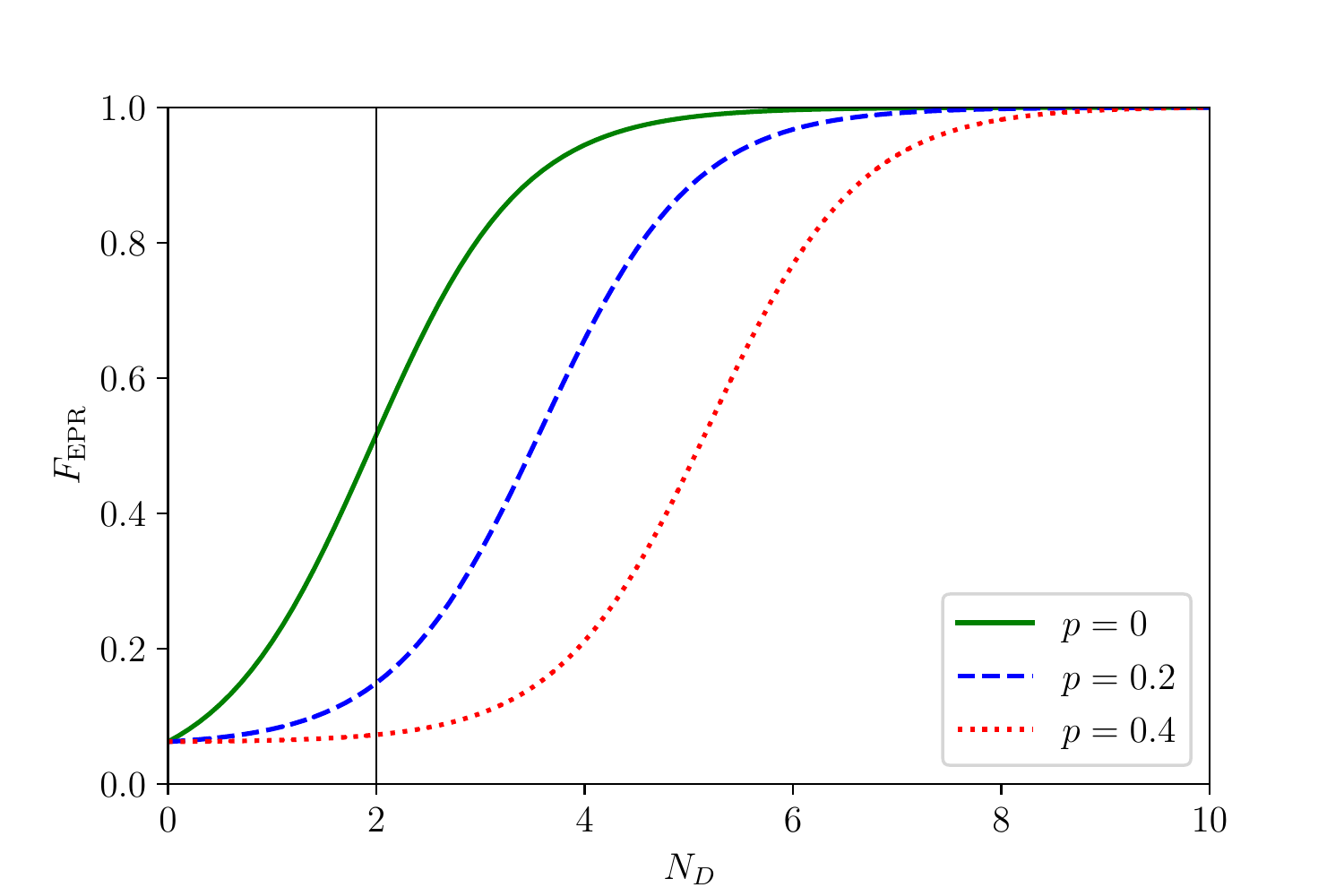}
\label{fig:Ferasure}
}\end{minipage}\
\begin{minipage}{.49\textwidth}
\subfloat[Decoherence]{
\includegraphics[scale=0.5]{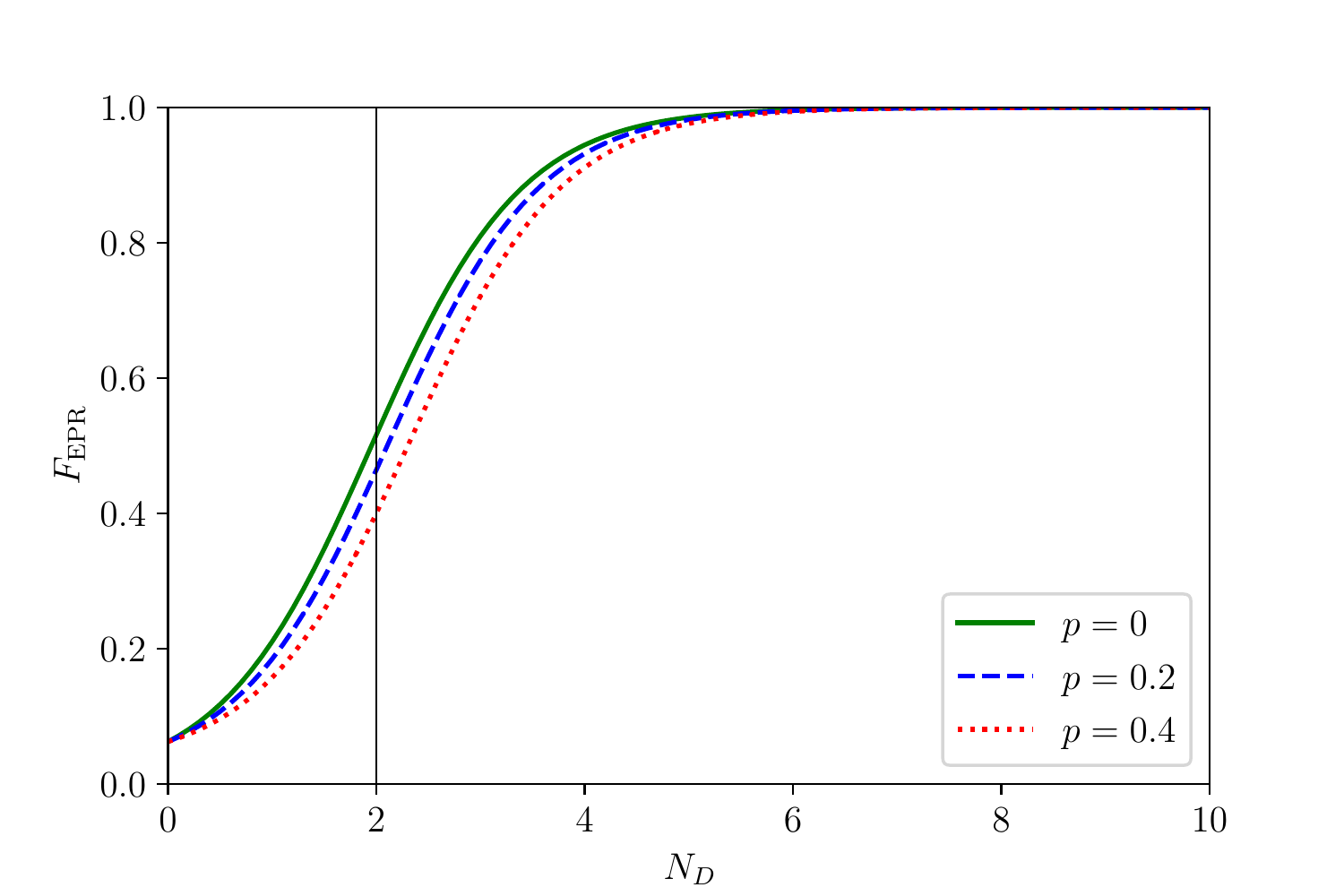}
\label{fig:Fdecoh}
}\end{minipage}
\caption{$N_D$ (the size of late radiation) dependence of the decoding fidelity in the presence of (a) erasure errors and (b) decoherence. The sizes of total system and Alice's state are fixed to be $N=10$ and $N_A=2$ (black vertical line), respectively.
The erasure errors have bigger impacts on the decoding fidelity/recoverability compared with the decoherence. Indeed, the decohrence has a small influence on the recoverability of the protocol.}
\label{fig:F}
\end{figure}

The decoding fidelity $\bar{F}_\text{EPR}$ also displays a sharp difference between the erasure and the decoherence effects, as shown in Figs.~\ref{fig:F_pdep} and \ref{fig:F}.
The decoding fidelity declines rapidly as the erasure probability $p$ increases, which significantly limits the recoverability of Hayden-Preskill thought experiment in the presence of erasure errors. In contrast, the recoverability is well protected against the decoherence effect as long as the size $N_D$ of the radiation that Bob collects later is large relative to that of Alice's state $N_A$.

%%%%%%%%%%%%%%%%%%%%%%%%%%%%%%%%%%%%%%%%%%%%%%%%%
\section{Conclusion}

Both erasure errors and decoherence are likely to happen in Bob's storage of radiation that he collected at an earlier time in the Hayden-Preskill thought experiment, and consequently, affect the quality of decoding algorithm. We have assessed their impact on the quantum recovery channel from the viewpoint of its decoding fidelity.
We found that serious breakdown of decoding fidelity occurs when the erasure error exists in the storage while the protocol is relatively resilient against the decoherence. This may equivalently be attributed to the reduction of the mutual information~\eqref{eq:Fepr_I_Erase} as a consequence of the decreased entanglement due to the lost qubits.

In the present work, we primarily focused on the noise and decoherence occurring during the storage phase motivated by the setup of the thought experiment. There is, however, a chance that these can affect other parts of circuit, something which has been addressed in~\cite{Yoshida:2018vly}. They explored these effects on the unitary evolution, and proposed an experimental implementation to diagnose information scrambling in quantum systems on a noisy device.

It is also interesting to see our recovery analysis in the presence of the erasure in relation to the entanglement phase transition~\cite{Li:2018zeno,Li2019:hybrid,Skinner:2019,Vasseur:2019,Zabalo:2020,Bao:2020phase,Choi:2020qec,Gullans:2020scalable}. In the latter context, projective measurements and random unitaries are repeatedly applied. Depending on the measurement rates the late time entanglement structure of a quantum system can be drastically changed if the rate at which the system is disentangled by the measurements overcomes the rate of random circuit entanglement generation. A crucial difference here, however, is that randomly chosen qubits are discarded in the erasure while the measurements results may be used to retain a certain amount of entanglement.

%%%%%%%%%%%%%%%%%%%%%%%%%%%%%%%%%%%%%%%%%%%%%%%%%

\section*{Acknowledgement}
We would like to thank Elizabeth Wildenhain and Aidan Chatwin-Davies for useful discussions. N.B. is supported by the National Science Foundation under grant number 82248-13067-44-PHPXH, by
the Department of Energy under grant number DE-SC0019380, and by the Computational
Science Initiative at Brookhaven National Laboratory.
Y.K. is supported by RIKEN-BNL Research Center.
%%%%%%%%%%%%%%%%%%%%%%%%%%%%%%%%%%%%%%%%%%%%%%%%%

%\newpage
\appendix

%%%%%%%%%%%%%%%%%%%%%%%%%%%%%%%%%%%%%%%%%%%%%%%%%
\section{Haar average}
\label{app:Haar_integral}

For an arbitrary function $f$ and a unitary operator $V$, the integral over the Haar measure satisfies the following properties:
\begin{align}
 &\int\diff U =1,
 \qquad
 \int\diff f(U) = \int\diff f(UV) = \int\diff f(VU).
\end{align}
Here are the Haar integral formulas useful for our computation:
\begin{align}
 \int\diff U\ U_{i_1j_1}U^\ast_{i_2j_2}
 &=\frac{\delta_{i_1i_2}\delta_{j_2j_1}}{d},
\\
 \int\diff U\  U_{i_1j_1}U_{i_2j_2}U^\ast_{i_3j_3}U^\ast_{i_4j_4}
 &=\frac{\delta_{i_1i_3}\delta_{i_2i_4}\delta_{j_1j_3}\delta_{j_2j_4} 
 + \delta_{i_1i_4}\delta_{i_2i_3}\delta_{j_1j_4}\delta_{j_2j_3}}{d^2-1}
 \nonumber\\
 &-\frac{\delta_{i_1i_3}\delta_{i_2i_4}\delta_{j_1j_4}\delta_{j_2j_3} 
 + \delta_{i_1i_4}\delta_{i_2i_3}\delta_{j_1j_3}\delta_{j_2j_4}}{d(d^2-1)},
\end{align}
where $d$ is the dimension of the Hilbert space that $U$ acts on.

We present the computations that are skipped in the main text.
The first one is the Haar integral for~\eqref{eq:PeprIdeal},
\begin{align}
\label{eq:barPepr_app}
\begin{split}
 &\int\diff U\  U_{a_1b_1c_1d_1}U^\ast_{a_2b_1c_2d_1}U_{a_2b_2c_2d_2}U^\ast_{a_1b_2c_1d_2}
 \\
 &=\frac{\delta_{a_1a_2}\delta_{b_1b_1}\delta_{c_1c_2}\delta_{d_1d_1}\ 
 \delta_{a_2a_1}\delta_{b_2b_2}\delta_{c_2c_1}\delta_{d_2d_2} 
 + \delta_{a_1a_1}\delta_{b_1b_2}\delta_{c_1c_1}\delta_{d_1d_2}\ 
 \delta_{a_2a_2}\delta_{b_1b_2}\delta_{c_2c_2}\delta_{d_1d_2} }{d^2-1}
 \\
  &-\frac{\delta_{a_1a_2}\delta_{b_1b_1}\delta_{c_1c_1}\delta_{d_1d_2}\ 
 \delta_{a_2a_1}\delta_{b_2b_2}\delta_{c_2c_2}\delta_{d_2d_1} 
 + \delta_{a_1a_1}\delta_{b_1b_2}\delta_{c_1c_2}\delta_{d_1d_1}\ 
 \delta_{a_2a_2}\delta_{b_1b_2}\delta_{c_2c_1}\delta_{d_2d_2}}{d(d^2-1)}
 \\
 &=\frac{d_Ad_B^2d_Cd_D^2 + d_A^2d_Bd_C^2d_D}{d^2-1}
 -\frac{d_Ad_B^2d_C^2d_D+ d_A^2d_Bd_Cd_D^2}{d(d^2-1)}
 \\
 &=\frac{d_A^2d_Bd_D}{d^2-1}\left[d_B^2 + d_C^2
 -\frac{d_C^2}{d_A^2}-1\right].
\end{split}
\end{align}

The second computation is used in~\eqref{eq:barDeltaErase},
\begin{align}
\label{eq:barDeltaErase_app}
\begin{split}
 &\int\diff U\  U_{a_1(b_1b_2)c_1d_1}U^\ast_{a_1(b_1b'_2)c_2d_1}U_{a_2(b'_1b'_2)c_2d_2}U^\ast_{a_2(b'_1b_2)c_1d_2}
 \\
 &=\frac{\delta_{a_1a_1}\delta_{b_1b_1}\delta_{b_2b'_2}\delta_{c_1c_2}\delta_{d_1d_1}\ 
 \delta_{a_2a_2}\delta_{b'_1b'_1}\delta_{b_2b'_2}\delta_{c_2c_1}\delta_{d_2d_2} 
 + \delta_{a_1a_2}\delta_{b_1b'_1}\delta_{b'_2b'_2}\delta_{c_1c_1}\delta_{d_1d_2}\ 
 \delta_{a_1a_2}\delta_{b_1b'_1}\delta_{b'_2b'_2}\delta_{c_2c_2}\delta_{d_1d_2} }{d^2-1}
 \\
  &-\frac{\delta_{a_1a_1}\delta_{b_1b_1}\delta_{b_2b'_2}\delta_{c_1c_1}\delta_{d_1d_2}\ 
 \delta_{a_2a_2}\delta_{b'_1b'_1}\delta_{b'_2b_2}\delta_{c_2c_2}\delta_{d_2d_1} 
 + \delta_{a_1a_2}\delta_{b_1b'_1}\delta_{b_2b_2}\delta_{c_1c_2}\delta_{d_1d_1}\ 
 \delta_{a_1a_2}\delta_{b_1b'_1}\delta_{b'_2b'_2}\delta_{c_2c_1}\delta_{d_2d_2}}{d(d^2-1)}
 \\
 &=\frac{d_A^2d_B^{2(1-p)}d_Cd_D^2 + d_Ad_Bd_C^2d_D}{d^2-1}
 -\frac{d_A^2d_B^{2(1-p)}d_C^2d_D+ d_Ad_Bd_Cd_D^2}{d(d^2-1)}
 \\
 &=\frac{d_Ad_Bd_D}{d^2-1}\left[\frac{d^2}{d_B^{2p}} + d_C^2
 -\frac{d_C^2}{d_B^{2p}}-1\right].
\end{split}
\end{align}

We next carry out the integral for~\eqref{eq:barPeprErase},
\begin{align}
\label{eq:barPeprErase_app}
\begin{split}
 &\int\diff U\  U_{a_1(b_1b_2)c_1d_1}U^\ast_{a_2(b_1b'_2)c_2d_1}U_{a_2(b'_1b'_2)c_2d_2}U^\ast_{a_1(b'_1b_2)c_1d_2}
 \\
 &=\frac{\delta_{a_1a_2}\delta_{b_1b_1}\delta_{b_2b'_2}\delta_{c_1c_2}\delta_{d_1d_1}\ 
 \delta_{a_2a_1}\delta_{b'_1b'_1}\delta_{b'_2b_2}\delta_{c_2c_1}\delta_{d_2d_2} 
 + \delta_{a_1a_1}\delta_{b_1b'_1}\delta_{b_2b_2}\delta_{c_1c_1}\delta_{d_1d_2}\ 
 \delta_{a_2a_2}\delta_{b_1b'_1}\delta_{b'_2b'_2}\delta_{c_2c_2}\delta_{d_1d_2} }{d^2-1}
 \\
  &-\frac{\delta_{a_1a_2}\delta_{b_1b_1}\delta_{b_2b'_2}\delta_{c_1c_1}\delta_{d_1d_2}\ 
 \delta_{a_2a_1}\delta_{b'_1b'_1}\delta_{b'_2b_2}\delta_{c_2c_2}\delta_{d_2d_1} 
 + \delta_{a_1a_1}\delta_{b_1b'_1}\delta_{b_2b_2}\delta_{c_1c_2}\delta_{d_1d_1}\ 
 \delta_{a_2a_2}\delta_{b_1b'_1}\delta_{b'_2b'_2}\delta_{c_2c_1}\delta_{d_2d_2}}{d(d^2-1)}
 \\
 &=\frac{d_Ad_B^{2(1-p)}d_Cd_D^2 + d_A^2d_Bd_C^2d_D}{d^2-1}
 -\frac{d_Ad_B^{2(1-p)}d_C^2d_D+ d_A^2d_Bd_Cd_D^2}{d(d^2-1)}
 \\
 &=\frac{d_A^2d_Bd_D}{d^2-1}
 \left[d_B^{2(1-p)} + d_C^2-\frac{d_C^2}{d_B^{2p}d_A^2}-1\right].
\end{split}
\end{align}

Finally, we carry out the following computation for \eqref{eq:barDeltaDecoh},
\begin{align}
\label{eq:barDeltaDecoh_app}
\begin{split}
 &\int\diff U\  U_{a_1b_1c_1d_1}U^\ast_{a_1b_2c_2d_1}U_{a_2b_2c_2d_2}U^\ast_{a_2b_1c_1d_2}
 \\
 &=\frac{\delta_{a_1a_1}\delta_{b_1b_2}\delta_{c_1c_2}\delta_{d_1d_1}\ 
 \delta_{a_2a_2}\delta_{b_2b_1}\delta_{c_2c_1}\delta_{d_2d_2} 
 + \delta_{a_1a_2}\delta_{b_1b_1}\delta_{c_1c_1}\delta_{d_1d_2}\ 
 \delta_{a_1a_2}\delta_{b_2b_2}\delta_{c_2c_2}\delta_{d_1d_2} }{d^2-1}
 \\
 &-\frac{\delta_{a_1a_1}\delta_{b_1b_2}\delta_{c_1c_1}\delta_{d_1d_2}\ 
 \delta_{a_2a_2}\delta_{b_2b_1}\delta_{c_2c_2}\delta_{d_2d_1} 
 + \delta_{a_1a_2}\delta_{b_1b_1}\delta_{c_1c_2}\delta_{d_1d_1}\ 
 \delta_{a_1a_2}\delta_{b_2b_2}\delta_{c_2c_1}\delta_{d_1d_1}}{d(d^2-1)}
 \\
 &=\frac{d_A^2d_Bd_Cd_D^2 + d_Ad_B^2d_C^2d_D}{d^2-1}
 -\frac{d_A^2d_Bd_C^2d_D+ d_Ad_B^2d_Cd_D^2}{d(d^2-1)}
  \\
 &=\frac{d_Ad_B^2d_D}{d^2-1}\left[d_A^2+d_C^2-\frac{d_A^2}{d_D^2}-1\right].
\end{split}
\end{align}
%%%%%%%%%%%%%%%%%%%%%%%%%%%%%%%%%%%%%%%%%%%%%%%%%
\section{Relation between $\calQ$ and $\tilde{\calQ}$}
\label{app:relation_decoh}

The quantum channels $Q$ and $\tilde{Q}$ are respectively defined by
\begin{align}
 \calQ(\rho) &= (1-p)\rho + p\frac{I}{d}\Tr\rho,
 \qquad
 \tilde{\calQ}(\rho) = (1-\tilde{p})\rho + \tilde{p}\frac{I}{d}\Tr\rho,
\end{align}
with $p = 2\tilde{p}-\tilde{p}^2$.

The identities used in \eqref{eq:rhoRBD2decoh} and \eqref{eq:rhoBD2decoh} are proved as follows:
\begin{align}
 &\figbox{0.4}{fig_tildeQ}
 =d_B\bra{\text{EPR}}_{B_2B_3} \tilde{Q}_{B_1B_2}\big(\ket{\text{EPR}}\bra{\text{EPR}}_{B_1B_2}\big)
 \otimes
 \tilde{Q}_{B_1B_2}\big(\ket{\text{EPR}}\bra{\text{EPR}}_{B_1B_2}\big)\ket{\text{EPR}}_{B_2B_3}
 \nonumber\\
 &=(1-\tilde{p})^2\ \figbox{0.4}{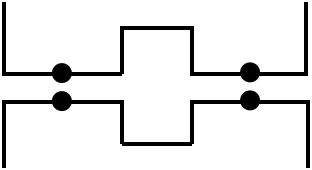}\ 
 +\ \frac{\tilde{p}(1-\tilde{p})}{d_B^2}\ \figbox{0.4}{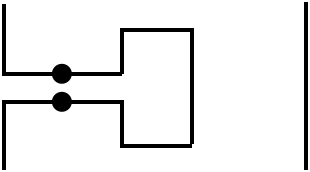}\ 
 +\ \frac{\tilde{p}(1-\tilde{p})}{d_B^2}\ \figbox{0.4}{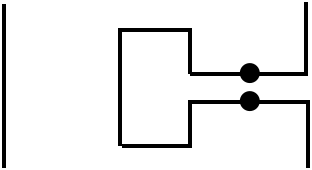}\ 
 +\ \frac{\tilde{p}^2}{d_B^4}\ \figbox{0.4}{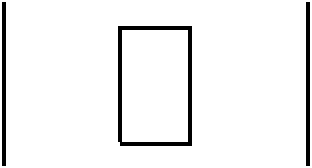}
 \nonumber\\
 &=\frac{1-(2\tilde{p}-\tilde{p}^2)}{d_B}\ket{\text{EPR}}\bra{\text{EPR}}_{B_1B_4}
 +\frac{2\tilde{p}-\tilde{p}^2}{d_B^3} I_{B_1B_4}
 \nonumber\\
 &= \frac{1}{d_B}Q_{B_1B_4}\big(\ket{\text{EPR}}\bra{\text{EPR}}_{B_1B_4}\big)
 =\figbox{0.4}{fig_tildeQ5}.
\end{align}
In the second to last equality, we used the identification $p=2\tilde{p}-\tilde{p}^2$.

%%%%%%%%%%%%%%%%%%%%%%%%%%%%%%%%%%%%%%%%%%%%%%%%%%
\bibliographystyle{utphys}
\bibliography{HPbib}

\end{document}